\documentclass[12pt]{article}  
\usepackage{a4,epsfig}
\begin{document}




\begin{titlepage}

\pagenumbering{arabic}
\vspace*{-1.5cm}
\begin{tabular*}{15.cm}{lc@{\extracolsep{\fill}}r}
{\bf Stockholms University} & 
\hspace*{1.3cm} \epsfig{figure=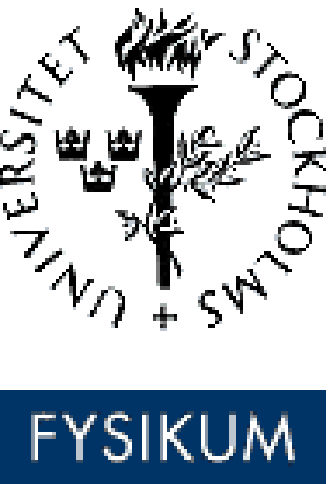,width=1.2cm,height=1.2cm}
&

USIP 2001-5
\\
& &

19  December, 2001
\\
&&\\ \hline
\end{tabular*}
\vspace*{2.cm}
\begin{center}
\Large 
{\bf \boldmath

Can SUSY be found at the Tevatron Run II ?
} \\
\vspace*{2.cm}
\normalsize { 

    {\bf A.~Lipniacka}

\par {\footnotesize Stockholm University, Stockholm
Center for Physics, Astronomy and Biotechnology, 
Fysikum, S - 106 91 Stockholm, Sweden}

}
\end{center}
\vspace{\fill}
\begin{abstract}
\noindent

Searches for SUSY particles are  the main physics focus of the ongoing
Run IIa  of the  Tevatron.  LEP results  have constrained  heavily the
Minimal  Supersymmetric Standard  Model.   In this  paper the  results
obtained  at  LEP and  the  Tevatron Run  I  are  analysed within  two
consistent  scenarios:  the gravity-mediated  MSSM  framework and  the
minimal  SUGRA scenario. In  these frameworks  limits much  beyond the
kinematic  reach  of LEP  are  set, and  the  allowed  mass range  for
particles which are not directly  observable at LEP is explored. These
results can be used to evaluate searches at the Tevatron Run II.  Both
R-parity   conserving   and   violating  scenarios   are   considered.
Model-dependence and coverage  of LEP results is discussed  in view of
searches at the Tevatron Run II,  and a conservative review of some of
the other constraints is given.  Consequences of the light Higgs boson
of  the  Minimal Supersymmetric  Standard  Model  with  mass close  to
114-117~GeV/c$^2$ and  branching ratio to $b  \bar{b}$ consistent with
the one expected in the Standard Model are examined.


\end{abstract}
\vspace{\fill}
\begin{center}
Presented at the Fourth Nordic Workshop for the LHC Physics, Stockholm 2001.
\end{center}
\vspace{\fill}
\end{titlepage}





%

\def\leqsim{\mathbin{\;\raise1pt\hbox{$<$}\kern-8pt\lower3pt\hbox{\small$\sim$}\;}}
\def\geqsim{\mathbin{\;\raise1pt\hbox{$>$}\kern-8pt\lower3pt\hbox{\small$\sim$}\;}}
\newcommand{\dfrac}[2]{\frac{\displaystyle #1}{\displaystyle #2}}
\def\MXN#1{\mbox{$ m_{\tilde{\chi}^0_#1}                                $}}
\def\MXNN#1#2{\mbox{$ m_{\tilde{\chi}^0_{#1,#2}}                        $}}
\def\MXNNN#1#2#3{\mbox{$ m_{\tilde{\chi}^0_{#1,#2,#3}}                  $}}
\def\MXC#1{\mbox{$ m_{\tilde{\chi}^{\pm}_#1}                            $}}
\def\XP#1{\mbox{$ \tilde{\chi}^+_#1                                     $}}
\def\XPP#1#2{\mbox{$ \tilde{\chi}^{+}_{#1,#2}                           $}}
\def\XCC#1#2{\mbox{$ \tilde{\chi}^{-}_{#1,#2}                             $}}
\def\XM#1{\mbox{$ \tilde{\chi}^-_#1                                     $}}
\def\XPM#1{\mbox{$ \tilde{\chi}^{\pm}_#1                                $}}
\def\XN#1{\mbox{$ \tilde{\chi}^0_#1                                     $}}
\def\XNN#1#2{\mbox{$ \tilde{\chi}^0_{#1,#2}                             $}}
\def\XNNN#1#2#3{\mbox{$ \tilde{\chi}^0_{#1,#2,#3}                       $}}
\def\p#1{\mbox{$ \mbox{\bf p}_1                                         $}}
\newcommand{\Gino}    {\mbox{$ \tilde{\mathrm G}                           $}}
\newcommand{\tanb}    {\mbox{$ \tan \beta                                  $}}
\newcommand{\smu}     {\mbox{$ \tilde{\mu}                                 $}}
\newcommand{\msmu}    {\mbox{$ m_{\tilde{\mu}}                             $}}
\newcommand{\msmur}   {\mbox{$ m_{\tilde{\mu}_R}                           $}}
\newcommand{\msmul}   {\mbox{$ m_{\tilde{\mu}_L}                           $}}
\newcommand{\sel}     {\mbox{$ \tilde{\mathrm e}                           $}}
\newcommand{\msel}    {\mbox{$ m_{\tilde{\mathrm e}}                       $}}
\newcommand{\stau}     {\mbox{$ \tilde{\tau}                               $}}
\newcommand{\stauo}     {\mbox{$ \tilde{\tau}_1                            $}}
\newcommand{\staut}     {\mbox{$ \tilde{\tau}_2                            $}}
\newcommand{\mstau}   {\mbox{$ m_{\tilde{\tau}}                            $}}
\newcommand{\mstauo}   {\mbox{$ m_{\tilde{\tau}_1}                         $}}
\newcommand{\mstaut}   {\mbox{$ m_{\tilde{\tau}_2}                         $}}
\newcommand{\snu}     {\mbox{$ \tilde\nu                                   $}}
\newcommand{\msnu}    {\mbox{$ m_{\tilde\nu}                               $}}
\newcommand{\msell}   {\mbox{$ m_{\tilde{\mathrm e}_L}                     $}}
\newcommand{\mselr}   {\mbox{$ m_{\tilde{\mathrm e}_R}                     $}}
\newcommand{\sell}   {\mbox{$ {\tilde{\mathrm e}_L}                     $}}
\newcommand{\selr}   {\mbox{$ {\tilde{\mathrm e}_R}                     $}}
\newcommand{\sfe}     {\mbox{$ \tilde{\mathrm f}                           $}}
\newcommand{\msfe}    {\mbox{$ m_{\tilde{\mathrm f}}                       $}}
\newcommand{\sle}     {\mbox{$ \tilde{\ell}                                $}}
\newcommand{\msle}    {\mbox{$ m_{\tilde{\ell}}                            $}}
\newcommand{\stq}     {\mbox{$ \tilde {\mathrm t}                          $}}
\newcommand{\mstq}    {\mbox{$ m_{\tilde {\mathrm t}}                      $}}
\newcommand{\sbq}     {\mbox{$ \tilde {\mathrm b}                          $}}
\newcommand{\msbq}    {\mbox{$ m_{\tilde {\mathrm b}}                      $}}
\newcommand{\msq}    {\mbox{$ m_{\tilde {\mathrm Q}}                      $}}
\newcommand{\An}      {\mbox{$ {\mathrm A}^0                               $}}
\newcommand{\hn}      {\mbox{$ {\mathrm h}^0                               $}}
\newcommand{\Zn}      {\mbox{$ {\mathrm Z}                                 $}}
\newcommand{\Zstar}   {\mbox{$ {\mathrm Z}^*                               $}}
\newcommand{\Hn}      {\mbox{$ {\mathrm H}^0                               $}}
\newcommand{\HP}      {\mbox{$ {\mathrm H}^+                               $}}
\newcommand{\HM}      {\mbox{$ {\mathrm H}^-                               $}}
\newcommand{\Wp}      {\mbox{$ {\mathrm W}^+                               $}}
\newcommand{\Wm}      {\mbox{$ {\mathrm W}^-                               $}}
\newcommand{\Wstar}   {\mbox{$ {\mathrm W}^*                               $}}
\newcommand{\WW}      {\mbox{$ {\mathrm W}^+{\mathrm W}^-                  $}}
\newcommand{\ZZ}      {\mbox{$ {\mathrm Z}{\mathrm Z}                      $}}
\newcommand{\HZ}      {\mbox{$ {\mathrm H}^0 {\mathrm Z}                   $}}
\newcommand{\GW}      {\mbox{$ \Gamma_{\mathrm W}                          $}}
\newcommand{\Zg}      {\mbox{$ \Zn \gamma                                  $}}
\newcommand{\sqs}     {\mbox{$ \sqrt{s}                                    $}}
\newcommand{\epm}     {\mbox{$ {\mathrm e}^{\pm}                           $}}
\newcommand{\ee}      {\mbox{$ {\mathrm e}^+ {\mathrm e}^-                 $}}
\newcommand{\mumu}    {\mbox{$ \mu^+ \mu^-                                 $}}
\newcommand{\eeto}    {\mbox{$ {\mathrm e}^+ {\mathrm e}^-\! \to\          $}}
\newcommand{\ellell}  {\mbox{$ \ell^+ \ell^-                               $}}
\newcommand{\eeWW}    {\mbox{$ \ee \rightarrow \WW                         $}}
\newcommand{\eV}      {\mbox{$ {\mathrm{eV}}                               $}}
\newcommand{\eVc}     {\mbox{$ {\mathrm{eV}}/c                             $}}
\newcommand{\eVcc}    {\mbox{$ {\mathrm{eV}}/c^2                           $}}
\newcommand{\MeV}     {\mbox{$ {\mathrm{MeV}}                              $}}
\newcommand{\MeVc}    {\mbox{$ {\mathrm{MeV}}/c                            $}}
\newcommand{\MeVcc}   {\mbox{$ {\mathrm{MeV}}/c^2                          $}}
\newcommand{\GeV}     {\mbox{$ {\mathrm{GeV}}                              $}}
\newcommand{\GeVc}    {\mbox{$ {\mathrm{GeV}}/c                            $}}
\newcommand{\GeVcc}   {\mbox{$ {\mathrm{GeV}}/c^2                          $}}
\newcommand{\TeV}     {\mbox{$ {\mathrm{TeV}}                              $}}
\newcommand{\TeVc}    {\mbox{$ {\mathrm{TeV}}/c                            $}}
\newcommand{\TeVcc}   {\mbox{$ {\mathrm{TeV}}/c^2                          $}}
\newcommand{\pbi}     {\mbox{$ {\mathrm{pb}}^{-1}                          $}}
\newcommand{\MZ}      {\mbox{$ m_{\mathrm Z}                               $}}
\newcommand{\MW}      {\mbox{$ m_{\mathrm W}                               $}}
\newcommand{\MA}      {\mbox{$ m_{\mathrm A}                               $}}
\newcommand{\GF}      {\mbox{$ {\mathrm G}_{\mathrm F}                     $}}
\newcommand{\MH}      {\mbox{$ m_{{\mathrm H}^0}                           $}}
\newcommand{\MHP}     {\mbox{$ m_{{\mathrm H}^\pm}                         $}}
\newcommand{\MSH}     {\mbox{$ m_{{\mathrm h}^0}                           $}}
\newcommand{\MT}      {\mbox{$ m_{\mathrm t}                               $}}
\newcommand{\GZ}      {\mbox{$ \Gamma_{{\mathrm Z} }                       $}}
\newcommand{\TT}      {\mbox{$ \mathrm T                                   $}}

\newcommand{\alphmz}  {\mbox{$ \alpha (m_{{\mathrm Z}})                    $}}
\newcommand{\alphas}  {\mbox{$ \alpha_{\mathrm s}                          $}}
\newcommand{\alphmsb} {\mbox{$ \alphas (m_{\mathrm Z})
                               _{\overline{\mathrm{MS}}}                   $}}
\newcommand{\alphbar} {\mbox{$ \overline{\alpha}_{\mathrm s}               $}}
\newcommand{\Ptau}    {\mbox{$ P_{\tau}                                    $}}
\newcommand{\mean}[1] {\mbox{$ \left\langle #1 \right\rangle               $}}
\newcommand{\dgree}   {\mbox{$ ^\circ                                      $}}
\newcommand{\qqg}     {\mbox{$ {\mathrm q}\bar{\mathrm q}\gamma            $}}
\newcommand{\Wev}     {\mbox{$ {\mathrm{W e}} \nu_{\mathrm e}              $}}
\newcommand{\Zvv}     {\mbox{$ \Zn \nu \bar{\nu}                           $}}
\newcommand{\Zee}     {\mbox{$ \Zn \ee                                     $}}
\newcommand{\ctw}     {\mbox{$ \cos\theta_{\mathrm W}                      $}}
\newcommand{\thw}     {\mbox{$ \theta_{\mathrm W}                          $}}
\newcommand{\thetabar}{\mbox{$ \theta^*                                    $}}
\newcommand{\phibar}  {\mbox{$ \phi^*                                      $}}
\newcommand{\thetapl} {\mbox{$ \theta_+                                    $}}
\newcommand{\phipl}   {\mbox{$ \phi_+                                      $}}
\newcommand{\thetamin}{\mbox{$ \theta_-                                    $}}
\newcommand{\phimin}  {\mbox{$ \phi_-                                      $}}
\newcommand{\ds}      {\mbox{$ {\mathrm d} \sigma                          $}}
\def    \ll           {\mbox{$\ell \ell                                    $}}
\def    \jjl          {\mbox{$jj \ell                           $}}
\def    \jj           {\mbox{$jj                                $}}
\def   \jjjj          {\mbox{${\it jets}                                   $}}
\newcommand{\jjlv}    {\mbox{$ j j \ell \nu                                $}}
\newcommand{\jjvv}    {\mbox{$ j j \nu \bar{\nu}                           $}}
\newcommand{\qqvv}    {\mbox{$ \mathrm{q \bar{q}} \nu \bar{\nu}            $}}
\newcommand{\qqll}    {\mbox{$ \mathrm{q \bar{q}} \ell \bar{\ell}          $}}
\newcommand{\jjll}    {\mbox{$ j j \ell \bar{\ell}                         $}}
\newcommand{\lvlv}    {\mbox{$ \ell \nu \ell \nu                           $}}
\newcommand{\dz}      {\mbox{$ \delta g_{\mathrm{W W Z}    }               $}}
\newcommand{\pT}      {\mbox{$ p_{\mathrm{T}}                              $}}
\newcommand{\ptr}     {\mbox{$ p_{\perp}                                   $}}
\newcommand{\ptrjet}  {\mbox{$ p_{\perp {\mathrm{jet}}}                    $}}
\newcommand{\Wvis}    {\mbox{$ {\mathrm W}_{\mathrm{vis}}                  $}}
\newcommand{\gamgam}  {\mbox{$ \gamma \gamma                               $}}
\newcommand{\qaqb}    {\mbox{$ {\mathrm q}_1 \bar{\mathrm q}_2             $}}
\newcommand{\qcqd}    {\mbox{$ {\mathrm q}_3 \bar{\mathrm q}_4             $}}
\newcommand{\bbbar}   {\mbox{$ {\mathrm b}\bar{\mathrm b}                  $}}
\newcommand{\ffbar}   {\mbox{$ {\mathrm f}\bar{\mathrm f}                  $}}
\newcommand{\ffbarp}  {\mbox{$ {\mathrm f}\bar{\mathrm f}'                 $}}
\newcommand{\qqbar}   {\mbox{$ {\mathrm q}\bar{\mathrm q}                  $}}
\newcommand{\nunubar} {\mbox{$ {\nu}\bar{\nu}                              $}}
\newcommand{\qqbarp}  {\mbox{$ {\mathrm q'}\bar{\mathrm q}'                $}}
\newcommand{\djoin}   {\mbox{$ d_{\mathrm{join}}                           $}}
\newcommand{\mErad}   {\mbox{$ \left\langle E_{\mathrm{rad}} \right\rangle $}}
\newcommand{\fpfbarp} {\mbox{$ {\mathrm f}'\bar{\mathrm f}'                $}}
\newcommand{\Lum}{${\cal L}\;$}
\newcommand{\lum}{{\cal L}}
\newcommand{\Cms}{$\mbox{ cm}^{-2} \mbox{ s}^{-1}\;$}
\newcommand{\cms}{\mbox{ cm}^{-2} \mbox{ s}^{-1}\;}
\newcommand{\Ecms}    {\mbox{$ E_{\mathrm{\small cms}}                      $}}
\newcommand{\Evis}    {\mbox{$ E_{\mathrm{\small vis}}                      $}}
\newcommand{\Erad}    {\mbox{$ E_{\mathrm{\small rad}}                      $}}
\newcommand{\Mvis}    {\mbox{$ m_{\mathrm{\small vis}}                      $}}
\newcommand{\pvis}    {\mbox{$ p_{\mathrm{\small vis}}                      $}}
\newcommand{\Minv}    {\mbox{$ m_{\mathrm{\small inv}}                      $}}
\newcommand{\pmiss}   {\mbox{$ p_{\mathrm{\small miss}}                     $}}
\newcommand{\ptmiss}  {\mbox{$ p_T^{\mathrm{\small miss}}                   $}}
\newcommand{\ptpair}  {\mbox{$ p_T^{\mathrm{\small pair}}                   $}}
\newcommand{\Mhfit}{\; \hat{m}_{H^0} }
\newcommand{\bl}      {\mbox{\ \ \ \ \ \ \ \ \ \ } }
\newcommand{\pL}      {\mbox{$ p_{\mathrm{L}}                              $}}
\newcommand{\Mrec}    {\mbox{$ m_{\mathrm{\small rec}}                      $}}
\newcommand{\Zto}   {\mbox{$\mathrm Z^0 \to$}}
\newcommand{\etal}  {\mbox{\it et al.}}
\def\NPB#1#2#3{{\rm Nucl.~Phys.} {\bf{B#1}} (19#2) #3}
\def\PLB#1#2#3{{\rm Phys.~Lett.} {\bf{B#1}} (19#2) #3}
\def\PLBN#1#2#3{{\rm Phys.~Lett.} {\bf{B#1}} (20#2) #3}
\def\PRD#1#2#3{{\rm Phys.~Rev.} {\bf{D#1}} (19#2) #3}
\def\PRL#1#2#3{{\rm Phys.~Rev.~Lett.} {\bf{#1}} (19#2) #3}
\def\ZPC#1#2#3{{\rm Z.~Phys.} {\bf C#1} (19#2) #3}
\def\EPJC#1#2#3{{\rm E.~Phys.~J.} {\bf C#1} (19#2) #3}
\def\PTP#1#2#3{{\rm Prog.~Theor.~Phys.} {\bf#1}  (19#2) #3}
\def\MPL#1#2#3{{\rm Mod.~Phys.~Lett.} {\bf#1} (19#2) #3}
\def\PR#1#2#3{{\rm Phys.~Rep.} {\bf#1} (19#2) #3}
\def\RMP#1#2#3{{\rm Rev.~Mod.~Phys.} {\bf#1} (19#2) #3}
\def\HPA#1#2#3{{\rm Helv.~Phys.~Acta} {\bf#1} (19#2) #3}
\def\NIMA#1#2#3{{\rm Nucl.~Instr.~and~Meth.} {\bf#1} (19#2) #3}
\def\CPC#1#2#3{{\rm Comp.~Phys.~Comm.} {\bf#1} (19#2) #3}

\def    \DM          {\mbox{$\Delta M$}}
\def    \missEt      {\ifmmode{/\mkern-11mu E_t}\else{${/\mkern-11mu E_t}$}\fi}
\def    \missE       {\ifmmode{/\mkern-11mu E}\else{${/\mkern-11mu E}$}\fi}
\def    \missp       {\ifmmode{/\mkern-11mu p}\else{${/\mkern-11mu p}$}\fi}
\def    \misspt      {\ifmmode{/\mkern-11mu p_t}\else{${/\mkern-11mu p_t}$}\fi}
\def    \DML         {\mbox{5~GeV $<\Delta M<$ 10~GeV}}
\def    \rs          {\mbox{$\sqrt{s}$}}
\def    \msneu       {\mbox{$m_{\tilde{\nu}}$}}

\section{Introduction \label{sec:intro}}

Supersymmetry (SUSY) is believed to be one of the most attractive scenarios
for physics beyond the Standard Model. In the last few years around 150 papers
on experimental searches for SUSY were
 published, out of which around 100 were related to the LEP
results and close to 30 to the  Tevatron results. This large number of papers
reflects perhaps as well the large number of free parameters relevant to
SUSY models at the presently explored energy scale.  
LEP and Tevatron are complementary from the experimental point of
view:  LEP
has lower energy reach, but  it is better suited to explore corners of 
SUSY models in a relatively assumption independent way, while the Tevatron  can
discover SUSY provided the Nature has chosen a version
of the model with favourable signatures.  For this reason
perhaps, LEP and Tevatron results are rarely analysed in a consistent
framework, and usually models  used to interpret
LEP results are less constrained.

In this paper the results obtained by LEP experiments  and these of 
the  Tevatron Run I 
are analysed within  two consistent scenarios: the gravity-mediated 
constrained MSSM framework and  the minimal SUGRA scenario. In
 these frameworks, limits
much beyond LEP's kinematic reach can be set, 
and the allowed mass range for particles which are not directly observable
at LEP (sneutrino and gluino) can be explored.
This has direct consequences for
designing the searches at the Tevatron.

In the  Minimal Supersymmetric extension of the Standard Model (MSSM) 
\cite{sugpeda},
each Standard Model particle has a supersymmetric partner with the same
couplings and with spin differing by 
$\hbar$/2.  Large corrections
to the Higgs mass from interactions involving  
virtual particles (heavy quarks in particular) are partially cancelled
due  to their superpartners. If they are lighter
than  1-10 \TeVcc\  this solves the so called
hierarchy problem \cite{SUSY2}. Moreover, supersymmetric particles
modify the  energy  dependence of the electromagnetic, weak and strong
coupling constants, and help them to unify at the scale of  
around $10^{15}$~\GeV  \cite{wim}.

The Higgs sector of the MSSM has to be extended to two complex
Higgs doublets $H_1,H_2$ responsible for  giving masses 
to the up and down-type
 fermions. 
Five physical Higgs boson mass states remain after the
Electroweak Symmetry breaking.
The lightest scalar neutral Higgs boson $\hn$ and the heavier pseudoscalar
neutral Higgs boson   $A$ are of interest for this paper.  
On the tree level, masses of the Higgs bosons depend on 
just two  parameters,
which can be chosen as  tan$\beta$, the ratio of vacuum expectation 
values of the two Higgs doublets,
and $m_A$. In particular   $m_h < m_Z*|cos2\beta| $ \footnote{For 
$m_A>>m_Z$, $m_{\hn(tree)} \sim  m_Z*|cos2\beta|/(1+m_Z^2/m_A^2) $, 
and for \tanb $\geqsim$ 10, $m_{\hn (tree)} \sim m_Z*|cos2\beta| $},  
however due to radiative
corrections mentioned above (which depend on the top quark mass, and 
on the mass terms of the superpartners of heavy quarks),
 the upper limit on the mass of  the lightest Higgs boson 
grows to  $m_h \leqsim$~135 \GeVcc\ \cite{higtherlim,carwig}.
                     
If $m_A \geqsim$ 150~\GeVcc\ the lightest supersymmetric Higgs boson
resembles very much the one of the Standard Model.
Precise  electroweak
measurements  \cite{lepew} 
suggest that the Higgs boson is relatively light
\footnote{The central value moves to $\sim$ 110~\GeVcc\ if
a different ansatz for hadronic corrections to the fine structure constant
is used}, $m_h = 88^{+53}_{-36}$\GeVcc, 
well in the range of the MSSM prediction. 
Searches for the Standard Model
like Higgs boson at LEP \cite{higdelphi,higlep} 
set a lower limit for $m_h$, 
$m_h>$114.1 \GeVcc\ ( if \tanb$<$ 6, or  $m_A>$120~\GeVcc),
constraining heavily the MSSM. The 2.1$\sigma$ ``excess'' observed at LEP 
\cite{fabiola} of events compatible with  
production of the Standard Model Higgs boson with   
$m_h \sim 114-117 $ \GeVcc,  together with the EW constraints, makes
low $m_h$, just above the reach of LEP, quite probable.
 The Run II of the Tevatron should cover the whole
mass range  allowed for  $m_{\hn}$ in the MSSM, providing  a definite
answer to whether the MSSM is a valid extension of 
the Standard Model \cite{higgsteva, ellihi}.  \\

The MSSM provides a phenomenologically interesting wealth of superpartners
of the Standard Model particles.
Supersymmetric partners of gauge 
and Higgs bosons (gauginos and higgsinos) 
mix to 
realize  four neutral mass states, neutralinos, $\XN{i}_{: i=1,4}$,
and four charged mass states, charginos, \XPM{1},\XPM{2}.
Superpartners of left-handed and right-handed fermions,
``right-handed'' and ``left-handed'' scalar quarks (squarks) and 
scalar leptons (sleptons) can mix. This leads to the off-diagonal
``left-right'' terms in their mass matrices and induces an 
additional  mass splitting
between the lighter and the heavier state.

While the Higgs sector is well constrained in the MSSM, very little
can be said about the superpartners mass spectrum unless one makes
some additional assumptions.
Experimental searches at LEP and the Tevatron (discussed 
in more detail in  section \ref{sec:limits})
constrain the lightest chargino and the sfermions to be heavier
than $\sim$ 100~\GeVcc, except for pathological mass configurations
which are  
discussed later. The Supersymmetry has thus  to be
broken. The pattern  of the  sparticle  mass spectrum 
depends primarily on
the mechanism of its  breaking. 

In the models with  gravity mediated supersymmetry breaking which
will be discussed in this paper,  the lightest neutralino (\XN{1}) is  
usually the
Lightest Supersymmetric Particle (LSP). If R-parity 
\footnote{$R$-parity is a multiplicative quantum number defined as 
$R=(-1)^{3(B-L)+2S}$
where $B$, $L$ and $S$ are the baryon number, the lepton number and the spin
of the particle, respectively. SM particles have R=+1 while 
their SUSY partners have $R=-1$}
is conserved the  LSP does not decay, and it is an ideal cold dark 
matter candidate
\cite{larsb}.   R-parity conservation was introduced 
to suppress baryon and lepton
number violating terms in the MSSM Lagrangian and thus 
to prevent the proton from decaying. However
it is not the only and perhaps not the best 
\cite{Ibanes} 
way to achieve this aim.
Constraints on models with broken R-parity will be 
only briefly discussed in this paper along
with a detailed discussion of R-parity conserving models. \\

Experimental searches motivated by the  MSSM with R-parity conservation 
and  gravity-mediated
supersymmetry breaking  exploit features of the model independent of 
further assumptions, like the 
strength of superpartner  couplings to  the gauge bosons, pair-production of
sparticles, and the missing energy and momentum signature due to escaping LSPs 
in the final state.
The same is to a large extent true for searches in the  MSSM 
with R-parity violation, 
except that single sparticle
production  and complicated decays of the LSP have to be taken into account.

However, to cover ``pathological'' situations with   final 
states which cannot be efficiently
detected  or situations  where the  production cross-sections are low, 
or finally to achieve more predictivity
and set limits on masses of the sparticles which are not directly 
observable (e.g. the LSP in the R-parity conserving model), 
additional model assumptions have to be made. In this paper 
two ``flavours'' of such
constraining assumptions are discussed (see section
\ref{sec:models}): the constrained MSSM with non-universal Higgs 
parameters (CMSSM with nUHP),
which is often used to interpret LEP results, and an even more constrained
minimal SUGRA scenario (mSUGRA) \footnote{ The definition of mSUGRA used
in this paper corresponds to what is called CMSSM with universal Higgs
masses in \protect{\cite{ellislight,benchellis,roszk}}}, 
often used to interpret 
Tevatron results  and
for  benchmark searches at  future colliders \cite{benchellis}.
It is shown in  section \ref{sec:limits}  that in both models  LEP results
can be used to exclude sparticles much beyond the kinematic limit of LEP.  
Perspectives to find sparticles at the Tevatron
are discussed in  section \ref{sec:tevatron}.

\section{ The models: CMSSM with nUHP and mSUGRA}
\label{sec:models}

To make the MSSM 
more predictive,  
the unification of some parameters at a high mass scale typical 
of Grand Unified Theories (GUT) can be assumed. 
In this section, approximate relations between the model parameters and
the superparners masses  which are important to 
understand the experimental limits
will be quoted without explanations. For a more complete 
information see  e.g. \cite{sugpeda}.

\noindent
\subsection{CMSSM with nUHP}
\label{sec:cmssm}

As well as  the already mentioned \tanb\ and $m_A$, the following parameters
are relevant in the constrained  MSSM with non-universal Higgs parameters:

\begin{itemize}

\item
\mbox{\boldmath$\mu$}, the Higgs mass parameter,

\item
\mbox{\boldmath $M_1,M_2,M_3$}, the  $U(1)\times SU(2)\times SU(3)$ 
gaugino masses 
at the electroweak (EW) scale.
Gaugino mass unification at the GUT scale is assumed, with a common gaugino 
mass
of $\boldmath m_{1/2}$. The  resulting relation between $M_1$ and $M_2$ is
$M_1=\frac{5}{3} tan^2\theta_W M_2\sim 0.5 M_2$,

\item
\mbox{\boldmath $m_{\tilde{\mathrm f}}$}, the sfermion masses.
Under the assumption of sfermion mass unification, \mbox{\boldmath${m_0}$} 
is the common sfermion mass at the GUT scale,

\item
the trilinear couplings \mbox{\boldmath$A_{\mathrm f}$} determining the 
mixing in the sfermion families.        
The third family trilinear couplings are the most relevant ones,   
\mbox{\boldmath $A_\tau, A_{\mathrm b}, A_{\mathrm t}$}.
\end{itemize}

Gaugino mass unification leads to  $m_{1/2} \simeq 1.2 M_2$ and 
to  the following approximate
relations between \MXC{1}, \MXN{1} and  the gluino mass ($m_{\tilde{g}}$):

\begin{itemize}
\item
in the  region where \XN{1} and \XPM{1} are {\bf gauginos}  ($|\mu| >> M_1$),
\MXC{1} $\simeq $ \MXN{2} $\simeq$ 2 \MXN{1},
 $ m_{\tilde{g}} \simeq$   3.2 \MXC{1} and
\MXC{1}  $\simeq$   $M_2$,
\item
in the {\bf higgsino} region ($|\mu| << M_1$), 
\MXC{1} $\simeq$ \MXN{2} $\simeq$  \MXN{1} $\simeq |\mu| $.\\ 
\end{itemize}

The relations between  chargino, neutralino and gluino masses and $|\mu|$
and $M_2$ are affected by radiative corrections of the order of 
2\%-20\% \cite{radcor}.
However, {\bf only the relative relations between 
chargino, neutralino and gluino
masses  are important from the experimental point of view}, 
and here the corrections
are much smaller. For example, the relation
\MXC{1}/\MXN{1} $\simeq$ 2 in the gaugino
region, which is usually
exploited to set a limit on the LSP mass, 
receives the corrections only of the order of  2\%;
and the ratio $m_{\tilde{g}}/\MXC{1}$ $\simeq$  3.2 receives corrections of the
order of 6\%. Thus, for example, the limit \cite{susywg} on the chargino mass
of 103.5~\GeVcc\  set by LEP 
(valid for \msnu$>$300~\GeVcc, \mstauo$>$\MXC{1}, 
and for $M_2 \leqsim 200$~\GeVcc) 
can be safely translated to \MXN{1} $\geqsim$ 51~\GeVcc\ and  
$m_{\tilde{g}} \geqsim$~310~\GeVcc.\\
   
If the  sleptons are heavy the chargino mass limit excludes regions 
in ($M_2,|\mu|$ ) plane  (see e.g. \cite{delphisusy}).
For  \tanb $\geqsim$, 2  $|\mu| \geqsim  $100 \GeVcc\ 
is excluded up to very
high values of $M_2$ (of the order of 1000~\GeVcc\ or more) while  
$M_2 \leqsim$ 100~\GeVcc\ is excluded for   
$|\mu| \geqsim  $100 \GeVcc. \\

Electroweak symmetry imposes the following relation between the masses of
the superpartners of the left-handed electron (\sell) and 
of the neutrino (\snu),

1) $\msell^2 = \msnu^2 + m_W^2 |cos2 \beta| $.

The assumption of sfermion mass unification relates masses
of the ``left-handed'' ($m_L$) and the ``right-handed'' 
($m_R$) ``light'' sfermions,  
``light'' squark  masses, and the gaugino mass
parameter  $M_2$. For example :

2)   $\msnu^2 $ $=$ $m_0^2+ 0.77M_2^2 -0.5\MZ ^2 |\cos 2\beta|$

3)   $m_L^2$ $=$ $m_0^2+ 0.77M_2^2  +(0.5 - sin^2\theta_W) \MZ ^2 |\cos 2\beta|$

4)   $m_R^2$ $=$ $m_0^2+ 0.22M_2^2 + sin^2\theta_W \MZ ^2 |\cos 2\beta|$
 
5)   $m_{d_L}$ $=$ $m_0^2 + 9 M_2^2 +(0.5 -1/3 sin^2\theta_W)\MZ ^2 |\cos 2\beta|$

Thus, for example,
   $m_{d_L} \geqsim$~310~\GeVcc, if \MXC{1} $\geqsim$ 103.5~\GeVcc.\\

Mixing between left and right states (present for superpartners of heavy
fermions) gives rise   to   off-diagonal  ``left-right'' mixing
terms in their mass matrices, which lead to a mass splitting between 
the lighter and the heavier state. 
 At the EW scale these terms are  proportional to
$m_{\tau}(A_\tau - \mu \tanb)$, $m_{b}(A_b - \mu \tanb)$ and
$m_{\mathrm t}(A_t - \mu / \tanb)$ for 
\stau, \sbq\ and \stq, respectively, where $A_{\tau}, A_b, A_t$ are free
parameters.
Therefore, for large $\mu$ this can give light stau and sbottom states
if \tanb\ is large, or a light stop for small \tanb.\\

For large $m_A$, the lightest Higgs boson mass depends primarily on  
\tanb,  $m_{top}$ and the mixing in the stop sector $X_t$
(expressed here as $X_t=A_t - \mu / \tanb $), and this  dependence
is maintained whether any additional constraints on the  MSSM
are imposed or not. The top quark mass is presently known with the
uncerntainty (1$\sigma$) of around 5~\GeVcc\ \cite{pdg}, and
the resulting uncerntainty of the lightest Higgs boson mass calculation
is around 6.5~\GeVcc, 
as $\Delta m_{\hn}/m_{\hn} \simeq 2 \Delta m_{top}/m_{top}$.
It was
shown in \cite{carwig} that for a given \tanb\ and top mass, the maximal
$m_{\hn}$ occurs for   $X_t/m_{SUSY}= \sqrt{6}$.  Another, slightly lower
maximum occurs for   $X_t/m_{SUSY}= -\sqrt{6}$. $m_{SUSY}$ is typically
taken to be of the order of the  gluino mass, or of the 
diagonal terms in the squark mass matrices, and $m_{\hn}$ grows 
with $m_{SUSY}$.

It should be noted that the off-diagonal terms in mass matrices of the third
family sparticles cannot be too big compared to the diagonal terms, in order
for a real solution for sparticle masses to exist. 
As diagonal terms grow  with  $m_0$ and $M_2$, for every
given value of the off-diagonal term 
a lower  limit is set on the corresponding 
combination of $m_0$ and $M_2$
\footnote{\protect{To avoid ``tachyonic'' mass solutions we must have:

$$m_{ll}+m_{rr}  > \sqrt{ (m_{ll}-m_{rr})^2 +4*m_{lr}^2 }$$

\noindent
where $m_{lr}$ is the off-diagonal mixing term, and $m_{ll},m_{rr}$
are the diagonal mass terms. For example, for the stop we have 
$m_{lr}=m_{top}X_t$ and,

$m_{ll} \simeq m_0^2 + 9M_2^2 +
m_{top}^2 + m_Z^2 cos2\beta (0.5-2/3sin^2\theta_W)$

$ m_{rr} \simeq m_0^2 + 8.3M_2^2 + m_{top}^2 + 
2/3 m_Z^2 cos2\beta sin^2\theta_W $

For an example value of $X_t=\sqrt{6}$ \TeVcc, 
the condition above sets a lower limit
on a combination of $m_0^2$  and   $M_2^2$:
${m_0}^2 + 8.5{M_2}^2 > 0.39$ \TeVcc\
Thus, if $m_0< 300$~\GeVcc\ we must have $M_2> 190$~\GeVcc. 
}}.

\noindent
\subsection{  {mSUGRA} }

In the minimal SUGRA model not only the sfermion masses, 
but also the Higgs masses 
$m_{H_1}$ and $m_{H_2}$, are assumed to unify to the common 
$m_0$ at the GUT scale. 
Then $m^2_{H_2}$ becomes negative at the  EW scale 
in most of the  parameter space, thus ensuring 
EW symmetry breaking. 

The additional requirements of the unification of the trilinear 
couplings to a common $A_0$
and the correct reproduction of the EW symmetry scale, which fixes the 
absolute value of $\mu$, defines the minimal gravity-broken MSSM
(mSUGRA).  The value   of $\mu^2$ can be determined minimising 
the Higgs potential and 
requiring  the right value of  $m_Z$. At tree level \cite{sugpeda}:
 
6) $\mu^2 =- 1/2m_Z^2+ \frac{ m^2_{H_1} -m^2_{H_2}tan^2\beta}{tan^2\beta -1}$

7)  $m^2_{H_1} \simeq  m^2_0 +0.5{m^2}_{1/2}$, 
$m^2_{H_2} \simeq -(0.275m^2_0+3.3{m^2}_{1/2})$

The parameter set is then reduced to  
$m_{1/2}, m_0, \tanb, A_0$ and the sign of $\mu$. 

In addition to the  mass relations listed in the previous subsection,  
$m_A$ can be related to $m_{1/2}$ ($M_2$), $m_0$ and Yukawa 
coupling of the top quark.
The stop
mixing parameter can be expressed (approximately) 
as $A_t =0.25 A_0 -2 m_{1/2}$ (\footnote{ For low \tanb,   
$m_A^2$ $\simeq$ $m_0^2 + 3m_{1/2}^2 -m_Z^2$. As $m_{\hn}$ grows
with $m_A$ and $A_t$ (see section \protect{\ref{sec:intro}}), 
Higgs searches  can be used to set a limit on  $m_{1/2}$ ($M_2$)
which depends on \tanb, $A_0$, and $m_{top}$}).
The
lightest Higgs mass can thus be related to 
$m_{1/2}$ ($M_2$), and the experimental
limit on it can be used to set  
limits 
 on the masses of 
(for example) the lightest
chargino and the lightest neutralino
dependent on
\tanb, $A_0$ and $m_{top}$.

\section{LEP and Tevatron results}
\label{sec:search}

In years  1995-2000, the Aleph, DELPHI, L3 and OPAL experiments at LEP 
collected an integrated
luminosity of more than  2000~\pbi\ at  centre-of-mass energies ranging 
from 130~GeV to 208~GeV.
These data have been analysed to search for the sfermions, 
charginos, neutralinos and Higgs bosons predicted by supersymmetric 
models 
\cite{higdelphi,delphisusy,opalsusy,l3susy,alephsusy,higaleph,higopal,higl3}.

Extensive searches for supersymmetry and Higgs were performed at the 
Run I of the
Tevatron with luminosity of around 100~\pbi\ \cite{rocco,sugtev}. 
LEP and Tevatron results used in this note
are discussed below.

\subsection{Searches for \hn\ production} 
\label{sec:higgs}

The Standard Model Higgs boson is produced  via the  Bjorken
``higgsstrahlung'' process, $\ee \to Z^* \to hZ$. 
Higgs searches (both in the  Standard Model and the MSSM) 
are  primarily sensitive to the  
Bjorken process with $h \to b\bar{b}$. Thus,  
the 95\% CL  lower limit
on the SM Higgs mass of $m_h > 114.1 $ \GeVcc\ set by LEP \cite{higlep}
corresponds approximately to the limit  
$\sigma(ee \to hZ)$ $ \times$ BR($h \to b \bar{b}$)$\leqsim$0.07~pb at 
$\sqrt{s}$=207~GeV.
This value can be used to set a conservative $\hn$ mass limit in the MSSM. 
It follows
that the SM mass limit holds  in the MSSM if  
\tanb$<$~6  and/or  $m_A>$120~\GeVcc\
(where $\ee \to Z^* \to \hn Z$ and  $\hn \to b\bar{b}$ dominate \footnote{
for smaller $m_A$ the production of $\hn A$ and $\hn \to c \bar{c}$
decays start to be important, and the experimental sensitivity
degrades}), unless
there are supersymmetric particles to which $\hn$ decays  or which enhance
other branching-ratios via virtual loops. 
For $\hn \to $ \XN{1} \XN{1}\  (or \hn\ decaying to   other
experimentally invisible final states) there exist a limit  
$m_h > 114.4 $ \GeVcc\
set by LEP \cite{higlep}.

For  $m_{A}$ $\le$ 1000~\GeVcc,
$A_{t}$-$\mu/\tanb$=$\sqrt{6}$~\TeVcc\ (the 
maximal  $m_{h^0}$ scenario  used in 
\protect{\cite{higlep}}), and $m_{top} =$ 174.3 \GeVcc,
the \tanb\ range,  0.5$\ge \tanb \ge$ 2.36, is excluded by the Higgs 
searches, if there are
no  supersymmetric particles which affect Higgs decay branching fractions
\cite{fabiola}.

The possible ``evidence'' observed at LEP for  a SM like Higgs boson with 
$m_h= 114-116$ \cite{fabiola} is 
based on a 2.1~$\sigma$ excess of $q \bar{q} b\bar{b}$ events compatible
with $\hn Z$ production. It can be translated into an approximate  
preferred region of
0.03~pb $\leqsim$ $\sigma(hZ)$ BR($h \to b \bar{b}$) $\leqsim$ 0.07~pb
at $\sqrt{s}$=207~GeV.

Searches at the Tevatron Run I \cite{rocco} impose 
$m_{\hn} > 120 $ \GeVcc\
for \tanb $>$ 70, and 
$m_{\hn} > 110 $ \GeVcc\
for \tanb $>$ 60.

\subsection{Searches for charginos and neutralinos  }
\label{sec:searchar}

After the  Higgs, charginos were the most important
SUSY discovery channel at LEP. Unless there is a light sneutrino 
\footnote{In the gaugino region
the chargino production cross-section can be quite small due to
the negative interference between the t-channel sneutrino exchange diagram
and the s-channel $Z/\gamma$ exchange diagram. 
Higgsino-type charginos do not
couple to the sneutrino.}, the  chargino pair production
cross-section is predicted to be large  if $\MXC{1}< \sqrt{s}/2$. 
A lower  limit
on the chargino mass of 103.5~\GeVcc\ was set
\cite{susywg}, assuming 100\% branching fraction
to  the R-parity conserving decay mode
 $\XPM{1} \to \XN{1} W^*$.
All  R-parity violating modes  were studied as well 
\cite{opalsusy,rdelphi,alephr},
and limits of 102-103~\GeVcc\ on the chargino mass were set. 
If the results of  all LEP experiments are combined, the limit  
will probably reach the kinematic 
limit of 104~\GeVcc. 
It should be noted that  if R-parity is violated, sneutrino decays lead
to visible final states; thus light sneutrinos (and the regions
in parameter space where the chargino production cross-section
is low) can be directly excluded from  the non-observation of 
sneutrino pair-production \cite{rdelphi,alephr}.

Cross-section limits for chargino pair-production 
were set. In R-parity conserving scenarios
they depend primarily on the difference between the mass of the chargino and an
undetectable sparticle it decays to (e.g. \XN{1} or \snu ).   
Chargino pair production with cross-section larger than  0.1-0.2~pb 
(corresponding  to 
$\sqrt{s} \sim$ 205 GeV, the average energy of the year 2000 data)
is  excluded 
 for $\DM~>~20~\GeVcc$ \cite{opalsusy,delamsb}, where
\DM~=~\MXC{1}~$-$~\MXN{1} or \DM~=~\MXC{1}~$-$~\msnu.
If these limits are combined, a  chargino production cross-section
above   0.05~pb-0.1~pb can be excluded.  The limit on chargino mass
of \MXC{1} $\geqsim$ 100~\GeVcc\ can be set  for the light
sneutrino as well,  as long as  $\DM~\geqsim~10~\GeVcc$.

If sfermion mass unification is assumed, searches for \selr\ can be used  
to set a lower limit on the sneutrino mass, and thus on the chargino mass
in the case of a light sneutrino and $\DM~<~10~\GeVcc$. Moreover,
if \sell\ and \selr\ are light, neutralino production in the
gaugino region is enhanced \footnote{Experimentally observable
neutralino production (for example \XN{1}\XN{2}) has quite
large cross-section in the higgsino region as higgsinos
couple directly to \Zn. However, in the gaugino region there is
no tree-level coupling of \XN{1} to \Zn, and $\ee \to \XN{1}\XN{2}$
can only be mediated via t-channel selectron exchange.}, and
neutralino searches 
set an indirect limit on the sneutrino mass in some regions of the parameter
space.

In R-parity conserving scenarios, another "blind-spot" in chargino 
searches arises when the \stauo\ is light and close in mass
to  the $\XN{1}$ \cite{delphisusy,susywg}. Chargino decays 
$\XPM{1} \to \stauo\ \nu$ with $\stauo \to \XN{1} \tau$ then
dominate, and lead to an ``invisible'' final state; but   
the search for neutralino production  can be used \cite{delphisusy,susywg} 
in this
case.
If neutralinos decay  via light stau states and 
$\mstau$ is close to $\MXN{1}$,
$\XN{1}\XN{2} $  production  with 
\XN{2}~$\to$~\stau$\tau$ and \stau~$\to$~\XN{1}$\tau$ leads to
only one $\tau$ visible in the detector;  nevertheless 
limits  on the cross-section times branching ratio 
are of the order of 0.1-0.4 pb \cite{delamsb}. 
The search for  $\XN{2}\XN{2}$ in the same region 
reaches  a sensitivity of 0.06 pb \cite{delphisusy}.  
In the CMSSM with nUHP, the region 
in ($M_2,\mu, m_0$) space where
the stau is degenerate in mass with the LSP depends on mixing parameters: 
$A_{\tau}$, 
and  $A_{b}$,$A_{t}$. It is possible to find  configurations of mixing
parameters  (typically
with $|\mu|$ few times larger than $M_2$ and $m_0$)  
such that the  stau is light and close in mass to \XN{1} while
the selectrons are heavy, rendering the neutralino cross-section small.
However, the chargino production cross-section is large in this case,
and this region can be 
explored  by the search for $\XPM{1}\XPM{1} \gamma$  production 
\cite{delphisusy,chadege,alephnote}  where the photon arises from 
initial state radiation  and is detected together with 
a few low energy
tracks originating from \XN{2}~$\to$~\stau$\tau$ and \stau~$\to$~\XN{1}$\tau$ 
decay chain.

In mSUGRA, $|\mu|^2$ is in the range
3.3~$m^2_{1/2}$-0.5$m_Z^2$~$ < \mu^2 < $~$m^2_{0}+3.8 m^2_{1/2}$ 
for \tanb$>$2 and
and light stau cannot be degenerate with neutralino
for large $m_0$. Thus  neutralino searches   
set  a limit on  the chargino mass  for small $\mstauo -\MXN{1}$ which 
is close to the one obtained for  heavy sleptons (around 103~\GeVcc). 

It is perhaps worth mentioning that, because  in the
higgsino region ($M_1>>|\mu|$)  the  $\XN{1} \XN{2}$ production
cross-sections at LEP are large,
$\XN{1} \XN{2}$ production can be excluded nearly up to the
kinematic limit as long as $\MXN{1}$ is not too close to $\MXN{2}$
($M_2 \leqsim 1500 $~\GeVcc\ in the constrained MSSM). 
For
$200 < M_2 < 1500 $~\GeVcc\ 
a lower limit on the LSP mass of 70~\GeVcc\ 
was set by DELPHI \cite{lim189},
using the data collected
at $\sqrt{s}$= 189~GeV. In the  constrained MSSM   
the mass difference between the 
lightest chargino and the lightest neutralino
is less  than  3~\GeVcc\ for $M_2 \geqsim 1500$~\GeVcc. 
A lower limit 
on the $\MXC{1}$ of around 86~\GeVcc\ was set in this region
by L3 Collaboration,\cite{l3susy}, 
implying a similar lower limit on the mass of the lightest
neutralino.   

\noindent
\subsection{Searches for Sleptons and Squarks}

Pair-produced selectrons and muons with the 
typical decay modes,
$\sle \to \XN{1} \ell$, have been searched for by all LEP
collaborations.
These searches exclude slepton pair production
with a cross-section above (0.02-0.1)~pb
depending on the
neutralino mass and on the slepton mass, 
assuming
100\% branching fraction to the above decay mode.
With this assumptions, right-handed smuons (selectrons) 
lighter than around 
96 (99)~\GeVcc\ can be excluded,
provided $\msmur(\mselr) -\MXN{1} \geqsim 20$~\GeVcc\ and that the
selectron pair production cross-section is as for \tanb=2, $\mu$=$-$200.
For the minimal coupling to $Z/\gamma$ and sufficiently
large  \DM~=~\mstauo~$-$~\MXN{1} $>$ 15 \GeVcc, 
\mstauo $\leqsim$ 85~\GeVcc\ can be excluded, while the lower limit 
on the mass of the stable stau is close to 87~\GeVcc.\\

The results of the LEP combined searches for  sbottom ($\sbq $) and
stop ($\stq $) 
are used in this paper.
The typical decay modes 
$\tilde{t} \to \XN{1}  c$ and $\tilde{b} \to \XN{1}  b$
have been searched for.
These searches exclude squark pair production
with a cross-section   above (0.05-0.1)~pb
depending on the
neutralino  and on the squark masses,
assuming
100\% branching fraction to the above decay modes. 
For the minimal coupling to
$Z/\gamma$  and for  
\DM~=~\mstq(\msbq)~$-$~\MXN{1} $>$ 15 \GeVcc, 
the \stq (\sbq) with  mass below
 95~(93)~\GeVcc\ is then excluded \cite{susywg}.

The CDF Run I searches exclude a stop quark lighter than 115~\GeVcc, 
if \MXN{1}$<$50~\GeVcc. However,  searches at LEP exclude \MXN{1}$<$50~\GeVcc\
in most of the parameter space (see section \ref{sec:limits}).
If CDF and LEP results on the sbottom searches are combined, a 
$\sbq_{1}$ lighter
than 140~\GeVcc\ can be excluded for \MXN{1}$<$60~\GeVcc\ and 
\DM~=~\msbq~$-$~\MXN{1} $>$ 7 \GeVcc\ \cite{susywg,sugtev}.

\section{Other experimental constraints}
\label{sec:other}

Other experimental  constraints which be  can used to evaluate SUSY models
are the measurements of the $b\to s \gamma$ decay rate (via
$B \to X_s \gamma$ decay), the measurement
of the  muon magnetic moment (g-2)  and, in R-parity  
conserving scenarios, the relict abundance of  
the LSP 
given by the model.  These constraints are discussed briefly 
below.\\

A possible  discrepancy
with the Standard Model was recently reported
in the g-2 experiment \cite{brook}, suggesting a presence
of sparticles \cite{benchellis,wimetc,others} lighter than a few hundrets
of \GeVcc's. However, the experimental results are still
very fresh and need a cross-check with more statistics.\\ 

In the constrained MSSM,
the SUSY contribution to $B \to X_s \gamma$ depends primarily on the
charged Higgs-top quark loops and 
chargino-stop quark loops. The result obtained by the
CLEO collaboration, 
$B \to X_s \gamma= (3.21\pm 0.43\pm 0.27^{+0.18}_{-0.10})10^{-4}$ \cite{CLEO},
is compatible with the recent result of BELLE 
$B \to X_s \gamma= (3.34\pm 0.50^{+0.34+0.26}_{-0.37-0.28})10^{-4}$ 
\cite{BELLE} 
and compatible with the next-to-leading order SM prediction
including  non-perturbative effects,  
$B \to X_s \gamma= (3.71\pm 0.31)10^{-4}$ \cite{misiak}.
Thus  the overall SUSY contribution (which can be either positive or negative)
has to be small. A relatively conservative estimate of the 
allowed range for the SUSY contribution would be to use  
weighted (with the statistical errors) average of BELLE and
CLEO results  and to treat systematic
errors and theory errors as fully correlated. The 
experimental result is then
$B \to X_s \gamma =(3.27\pm 1.15 (2.5 \sigma) \pm 0.27 \pm 0.2)10^{-4}$.
Using $2.5~\sigma$ (95`$\%$
confidence level) statistical error and assuming that the
systematic and theory errors can induce a correlated
shift of the result, 
one arrives at a conservative allowed
range for  the discrepancy beetween the theoretical (SM+MSSM) and the
experimental value (EX); 
$ -2 \times 10^{-4} \leqsim$~(MSSM+SM)-EX~$\leqsim 2 \times 10^{-4}$.

In the constrained MSSM,  SUSY contribution
 is small for   a large 
$m_A$ (and thus large $m_{H^\pm}$ as
$M^2_{H^\pm }$=$M^2_W + M^2_A$ ), and in  
the gaugino region for the chargino  where
the chargino-stop coupling  is smaller.  For example,
the charged 
Higgs-top contribution is below
$0.25(1) \times 10^{-4}$ for  
$m_A> 1000 (250)$~\GeVcc\  and low \tanb\ \cite{pokorstbg,carenam}. 
The chargino-stop
contribution in the gaugino region ($M_2 <0.2 |\mu| $) is of the order
of $0.25 \times 10^{-4}$ for $\MXC{1}=\mstq=300$~\GeVcc\ and of the
order of $2 \times 10^{-4}$ for $\MXC{1}=\mstq=100$~\GeVcc,  close to the
experimental lower mass limit.
The charged Higgs-top contribution
always adds  to the SM one, whereas there can be a destructive interference
between the SM and the chargino-stop contribution. 
Both contributions grow at large \tanb.
The charged Higgs-top contribution contains \tanb\
dependent NLO terms.  The chargino-stop  
contribution  grows with    $ \sim A_t\mu\tanb $  for high values
of \tanb\ 
already in  the leading order.
It was shown in \cite{carenam} that  there is
a cancelation between  the leading and next-to-leading order terms 
 for $A_t>0$ and $\mu>0$  and
the chargino-stop contribution is never very large. For example, for 
$A_t=\mu=500~\GeV$, $\mstq=250$~\GeVcc\ and  \tanb$<$40
 the chargino-stop contribution
is smaller than 0.25$\times 10^{-4}$. 
For  $A_t=-\mu=-500~\GeV$ and \tanb=20(40) it is however of the order
of 2(4)  and, if $A_t\mu <0$,   can be partially cancelled
by the charged Higgs-stop contribution. 
Thus  if $A_t<0$, the
positive $\mu$ is favoured  and for $\tanb>$20 one must have 
either $\stq_{1}$ or $\XPM{1}$ heavier than $ \sim $250~\GeVcc\ or 
$m_{H^{\pm}} (m_A)$ of the order of 250(200)~\GeVcc. 

In mSUGRA  $A_t \simeq 0.25 A_0 - 2 m_{1/2}$, thus
it  is negative unless one considers large and positive $A_0$ values.
As $M_2, \mu$, $m_{H^{\pm}}$, $A_t$, and  the stop
and the chargino masses are  related, the $B \to X_s \gamma$ measurements
can be used to exclude regions of $(m_{1/2}(M_2);m_0)$ space. 
This was discussed
in e.g. \cite{benchellis,roszk} for $A_0=0$.
Somewhat less  conservative estimate of the allowed experimental range
of $B \to X_s \gamma$ was used there  than sugested in this paper.   
At $\tanb<10$, there is essentially
no constraints for $\mu>0$ and $\MXC{1} \geqsim 100$~\GeVcc,
whereas either a heavy
chargino or a squark is required if  $\mu<0$. For $\tanb> 20 (35)$, the limit
of $\sim 200(300)$~\GeVcc\ is set for $m_{1/2}$ if $m_0 \leqsim 600$~\GeVcc.
This corresponds to a lower limit on the diagonal terms in the 
squark mass matrix (or
on the masses of the superpartners of light quarks)
of the order of 600-700~\GeVcc. However,  the sbottom squark
can still be made lighter via mixing. 
The bounds mentioned above cannot have an interpretation
of limits at 95\% level. For that, a more sophisticated estimate
of the allowed range of the experimental value should be employed,
and  SUSY
contribution to    $B \to X_s \gamma$ 
should be calculated for  $A_0$ values other than 0 (\footnote{ Such
an analysis, employing $B \to X_s \gamma$
g-2, and Higgs constraints  was perfomed in \protect{\cite{wimetc}}. 
The conclusions will change however if present CLEO \cite{CLEO}
results, which bring the experimental average closer
to the SM result, are used.})\\ 

If  R-parity is conserved, the Universe could be filled
with the neutralino relict  of the Big Bang. The relict 
LSP density ($\Omega_{LSP} h^2$)
is governed by their annihilation rate
at  decoupling time \cite{larsb}. If the annihilation rate
was too small we could have enough LSP Dark Matter to
have collapsed the Universe by now.  The annihilation rate is  
proportional to the neutralino self-interaction
cross-section and   to the interaction cross-section
between LSPs and other supersymmetric particles (co-annihilation)
which are suitably close in mass (see e.g. \cite{edsjo}). 

For higgsino type  neutralinos, the self-interaction can be mediated by
the $\Zn$ and the cross-sections are large enough to keep
$\Omega_{LSP} h^2<$~0.3, as long as \MXN{1}~$\leqsim$~10~\TeVcc.\\

For gaugino-like neutralinos the cross-sections
are smaller, and the tree-level interaction  process has to be mediated 
by sfermions
or Higgs bosons. The annihilation rate is roughly inversely
proportional to the neutralino and the slepton mass scale, if
one neglects the 
resonant annihilation $\XN{1} \XN{1} \to H$ \cite{benchellis}
or $\XN{1} \XN{1} \to A$ \cite{roszk} or the resonant 
co-annihilation $\XN{1} \stauo$ \cite{benchellis}.
Except for $\XN{1} \XN{1} \to A (H)$, which can occur for high $m_{A(H)}$ 
 when
the width of $A(H)$ is large,  other resonant annihilation
conditions imply a slightly fine tuned mass relation between
\XN{1} and \stauo. 

If one ignores the resonant annihilation
channels mentioned above,  $\Omega_{LSP} h^2 <$~0.3  imposes 
$m_0 \leqsim $500~\GeVcc\ and $M_2 \leqsim 600$~\GeVcc\
for \tanb$<$50, and 
$m_0 \leqsim $150(200)~\GeVcc\ and $M_2 \leqsim 300$~\GeVcc\
for \tanb$<$10(20). This implies an upper limit on the LSP 
mass of 150-300~\GeVcc\ depending on \tanb\, and, for example, 
on  the \mselr\ mass of 200-600~\GeVcc, 
if the lightest neutralino is a gaugino.
This is typically the case in mSUGRA, where 
3.3~$m^2_{1/2}$-0.5$m_Z^2$~$ < \mu^2 $, 
and the gaugino region is thus favoured.

Neutralino dark matter  has received a lot of attention in the literature
(see for example \cite{roszk} and references therein). To set constraints
on SUSY models it is often required that the  LSP relict density should
provide all the non-baryonic dark matter for which  experimental evidence
exists. This implies a lower bound on  $\Omega_{LSP} h^2$. 
However, such a bound cannot be regarded as an experimental constraint,
as there may be other suitable sources of the non-baryonic dark
matter.  

\section{Limits}
\label{sec:limits}                                           

The searches described in the previous section were used to set limits
on sparticles masses in the CMSSM with non universal Higgs parameters and 
in mSUGRA. Whenever available, combined LEP cross-section limits and 
mass limits were used. These concern chargino, slepton and squark searches.
For the neutralino cascade decays via stau which were searched for so far
only by DELPHI and ALEPH 
\cite{delphisusy,alephnote,lim189,lim200}, 
it was assumed that other LEP collaborations
can reach a similar sensitivity. 
It was also assumed that $\sigma(hZ)$~BR($h \to b \bar{b}$)$\leqsim$0.07~pb, 
in accordance with the results of searches for the Higgs boson production. 
$m_{top}=174.3$~\GeVcc\ was used, the dependence of results on this 
value is discussed further.

{\bf Limits presented in this section are valid in the R-parity conserving
scenario and in all R-parity violating scenarios where a  chargino limit
of 103~{\boldmath \GeVcc }
or more can be set by LEP experiments, as discussed below.}

\subsection{Limits in the CMSSM with nUHP}

Higgs boson searches and chargino searches set limits in this scenario. 
"Holes" which arise  in chargino searches in the 
R-parity conserving scenario
are covered
by selectron, neutralino, Higgs and squark searches. All limits presented
in this section are for $m_A$=1000~\GeVcc. This choice is conservative
from the point of view of the experimental limit
set at low and moderate \tanb\ becouse 
the $\hn $ mass  grows with $m_A$.  Although for 
$92 <m_A<120$~\GeVcc\
and \tanb$>$6  the limit on $m_{\hn}$ degrades to 91-110~\GeVcc,
this is not expected to affect significantly  
any of the results  presented
in this section.   High value of $m_A$ ensures that the SUSY
contribution to $b \to s \gamma$ is small, at least
for $A_t>0$, in agreement with
the present experimental value 
\cite{CLEO,BELLE,pokorstbg,carenam}(see
section \ref{sec:other}).\\

The following range of parameters was studied; 
 $-$2000~$\le \mu \le$~2000~\GeVcc,
 0~$\le M_2 \le$~2000~\GeVcc,  0~$\le m_0 \le$~1000~\GeVcc,
 3~$\le \tanb\ \le$~60, 0~$\le A_t  \le$~500~\GeVcc, $A_{\tau}=A_{b}=0$.\\

\noindent
\underline{Limits on the  mass of the lightest neutralino}

\begin{figure}
\begin{center}
\vskip 0.1 cm
\mbox{\epsfysize=14.0cm\epsfxsize=14cm\epsffile{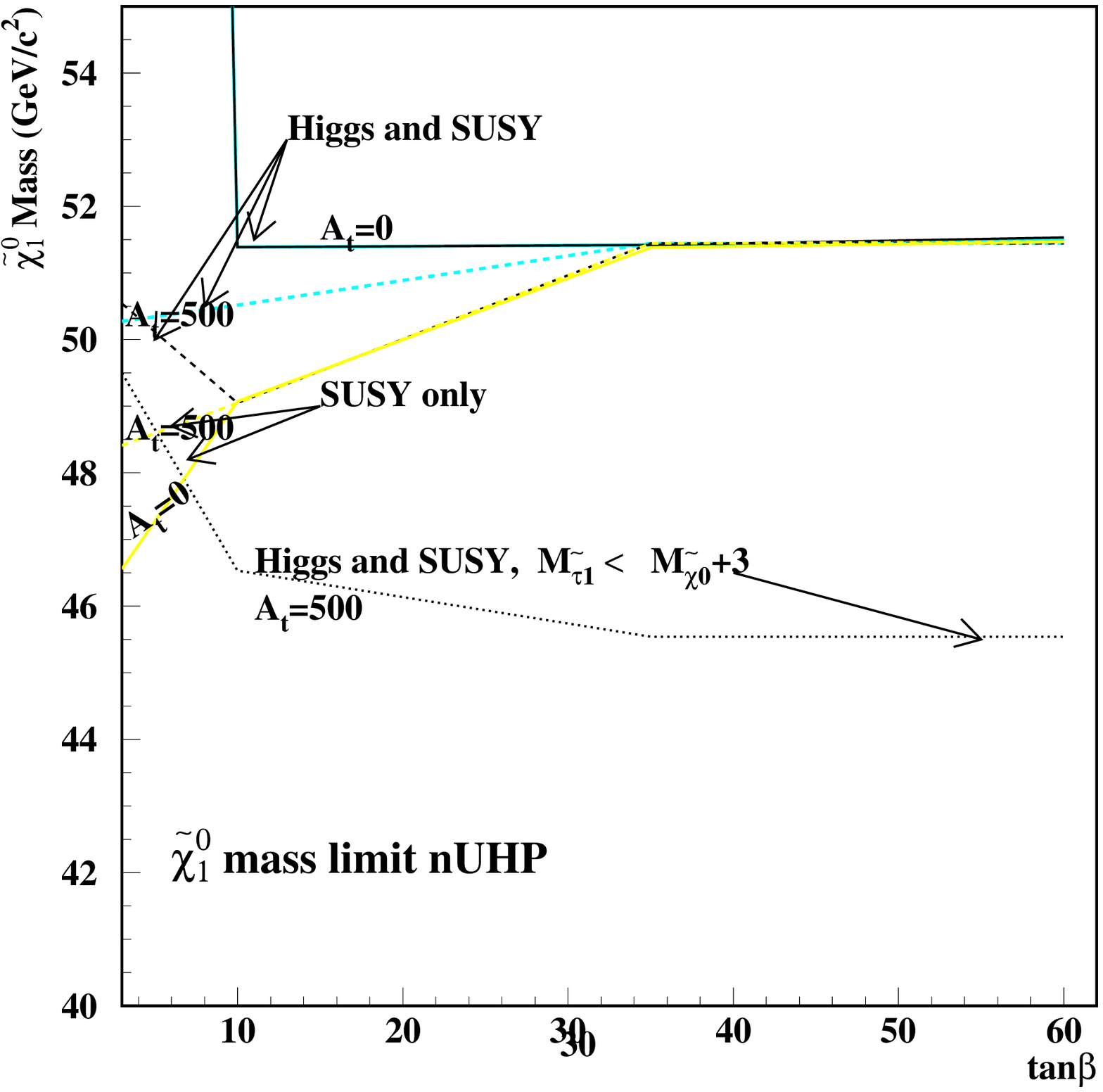}}
\caption[MSSM limits in ($\mu$,$M_2$) plane]{
The lower limit at 95~\% confidence level on the mass of the lightest
neutralino, \XN{1}, as a function of \tanb\ assuming a stable \XN{1}.
The dashed  and dotted (solid) curves shows limits obtained for 
$A_t$~=500~\GeVcc\ ($A_t$~=0~\GeVcc), and with mixing in the 
third family of the form:
($A_{\tau}-\mu\tanb$, $A_{b}-\mu\tanb$, $A_{t}-\mu/\tanb$, with
$A_{b}=A_{\tau}=0$).
For lines marked ``Higgs and SUSY'', constraints both from SUSY
and Higgs searches were imposed. Thin dark lines show the limit
obtained when a lower limit on the Higgs production cross-section
as described in the text was imposed, while  
for the limit shown with the thicker lighter line, it was assumed 
that 2.1~$\sigma$ ``excess'' observed by LEP represents a real
signal and it  
was required that
0.03~pb $\leqsim$ $\sigma(hZ)$ BR($h \to b \bar{b}$) $\leqsim$ 0.07~pb at
$\sqrt{s}=207$~GeV.
This condition excludes regions where  $\hn $ decays to supersymmetric
particles or is heavier than 117 GeV.
For $A_t=0$, thicker and thinner lines coincide.
The dotted line shows the limit in the region where \mstauo=\MXN{1}.
For other limits a condition \mstauo-\MXN{1}$>$5~\GeVcc\ was imposed.  
"SUSY only" lines drop at low \tanb\  due to a "hole"
in chargino searches when the  chargino is close in mass to 
the sneutrino.  This   hole is partially covered by
selectron and neutralino searches. See the text for more information
}
\label{fig:lspcmssm}
\end{center}
\end{figure}

The effect of various searches in the R-parity conserving scenario
is illustrated on figure \ref{fig:lspcmssm} showing the LSP mass
limit set by the  Higgs and SUSY 
searches and by the  SUSY searches alone, as a function of \tanb. 
The mixing in the third family was of the form:
($A_{\tau}-\mu\tanb$, $A_{b}-\mu\tanb$, $A_{t}-\mu/\tanb$), and plots are
 shown for two values of $A_t$.
The  LSP mass limit ranges from 
46-51~\GeVcc\ depending on the scenario, as discussed below.

If only the SUSY searches are exploited the limit drops at \tanb$<$10 due
to the "hole" in chargino searches, where the chargino is close in mass
to the sneutrino. The "hole" is partially covered by selectron 
and neutralino
searches, and it is less "deep" for
$A_t$=500~\GeVcc\ as higher
$m_0$ is required to avoid  the tachyonic stop (section \ref{sec:cmssm}).
It is covered by the Higgs search for $A_t=0$, as higher $M_2$ and
$m_0$ are required to get $m_{\hn} \geqsim$ 114~\GeVcc. 
At higher \tanb\ values this sneutrino  hole  is covered, because
a higher $m_0$ is required to get  the $\stauo$ heavier than the 
experimental limit, as the mass splitting between the heavier 
and the lighter
stau grows with $|A_{\tau}-\mu\tanb|$.
If \mstauo=\MXN{1} is allowed (the dotted line) the limit drops
at high \tanb\ to 46~\GeVcc, because  another  hole  in chargino and 
stau searches develops.
This "hole" is partially covered by neutralino and ``degenerate'' 
chargino searches, but
it is not covered by the Higgs searches (unless it is assumed
that 2.1~$\sigma$ observed at LEP represents the real signal
and  
0.03~pb~$\leqsim$ $\sigma(\hn Z)$ BR($\hn \to b \bar{b}$) $\leqsim$ 0.07~pb
is imposed, which  excludes
 $\hn \to \stauo \stauo$ which dominates in this region).
This limit represents the most
conservative scenario and it  is maintained even if
 mixing in  all the three families is
treated as totally independent and assumed to have an arbitrary $\mu$ dependence.
For other limits the condition \mstauo-\MXN{1}$>$5~\GeVcc\ was imposed. 
Radiative corrections according to \cite{radcor} were applied.

Only the range  of $\tanb >3$ was analysed, because  in the mixing model
used in this paper $\tanb <3$ is excluded by the Higgs search. Although
$2.4 (2.0) \leqsim \tanb <3$ for $m_{top}=174.3 (179) $ \GeVcc\
is allowed for the maximal mixing in the
stop sector, a relatively high $m_0$ or $M_2$ 
is implied either to avoid a tachyonic stop,
or to obtain $m_{\hn} \geqsim$ 114~\GeVcc. 

Thus, the limit $\MXN{1}> 46$~\GeVcc\
is valid for  $\tanb>1$, and essentially independent of 
the mixing scenario.\\

If R-parity is violated there are no  holes in chargino searches.
Both for purely leptonic and purely hadronic R-parity violating
terms, the limit $\MXN{1}>$ 50-51~\GeVcc\ was set by DELPHI \cite{rdelphi} 
for $\tanb> 2$. 

For a  mixed leptonic-hadronic (LQ$\bar{D}$) R-parity
violating term, ALEPH's 
limit on the chargino mass at high $m_0$ can be
translated to  \MXN{1} $\geqsim$ 51~\GeVcc\ at $\tanb>2$. 
Although the region of the lowest chargino production
cross-section  
is excluded by the  $\snu_e$  and \selr\  mass 
limits of  91 and 93~\GeVcc\  \cite{alephr}   
and by the Higgs limit, 
the cross-section limits from all LEP
experiments have to be combined to  
set a similar mass limit on the lightest chargino  
(and neutralino) as for  purely hadronic and leptonic terms.\\

The neutralino mass limit is set in the {\bf gaugino} region, at high $|\mu|$ 
values. In the {\bf higgsino} region  $\XN{1} \XN{2}$ and 
$\XPM{1} \XPM{1}$ production
cross-sections at LEP are higher 
and $\MXN{1}$ is closer to $\MXC{1}$, than in the
gaugino region. In the higgsino region $\MXC{1} < 86$~\GeVcc\ can be excluded
even if $\MXN{1}$ is very close or equal to $\MXC{1}$ \cite{l3susy}.
For $|\mu| <$~0.5~$M_2$, limits
on the cross-sections for  $\XN{1} \XN{2}$ and $\XPM{1} \XPM{1}$
production set by LEP exclude $\MXN{1}<$~80~\GeVcc. \\

\noindent
\underline{Limits on the sfermion masses and on the gluino mass}

\begin{figure}
\begin{center}
\vskip -0.5 cm
\mbox{\epsfysize=14.0cm\epsfxsize=14cm\epsffile{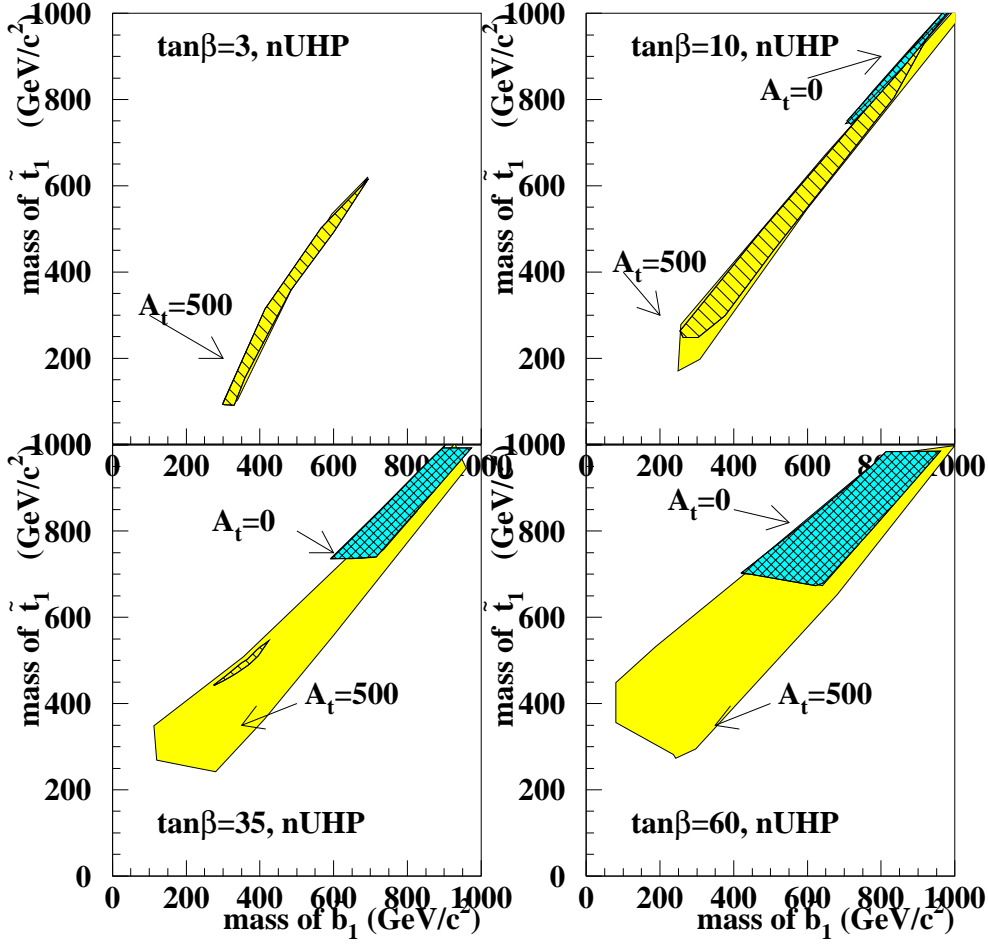}}
\caption[MSSM limits in ($\mu$,$M_2$) plane]{
Allowed range in  ($m_{\stq_{1}}$, $m_{\sbq_{1}}$)
plane at 95~\% confidence
for several values of \tanb, resulting from Higgs 
and SUSY searches.  The darker (lighter)  region is
for $A_t=0$ ($A_t$=500~\GeVcc). The hatched (cross-hatched)
region on the lighter (darker) shading is allowed for
0.03~pb $\leqsim$ $\sigma(\hn Z)$BR($\hn \to b \bar{b}$) $\leqsim$ 0.07~pb.
}
\label{fig:squark}
\end{center}
\end{figure}

\begin{figure}
\begin{center}
\vskip -0.5 cm
\mbox{\epsfysize=14.0cm\epsfxsize=14cm\epsffile{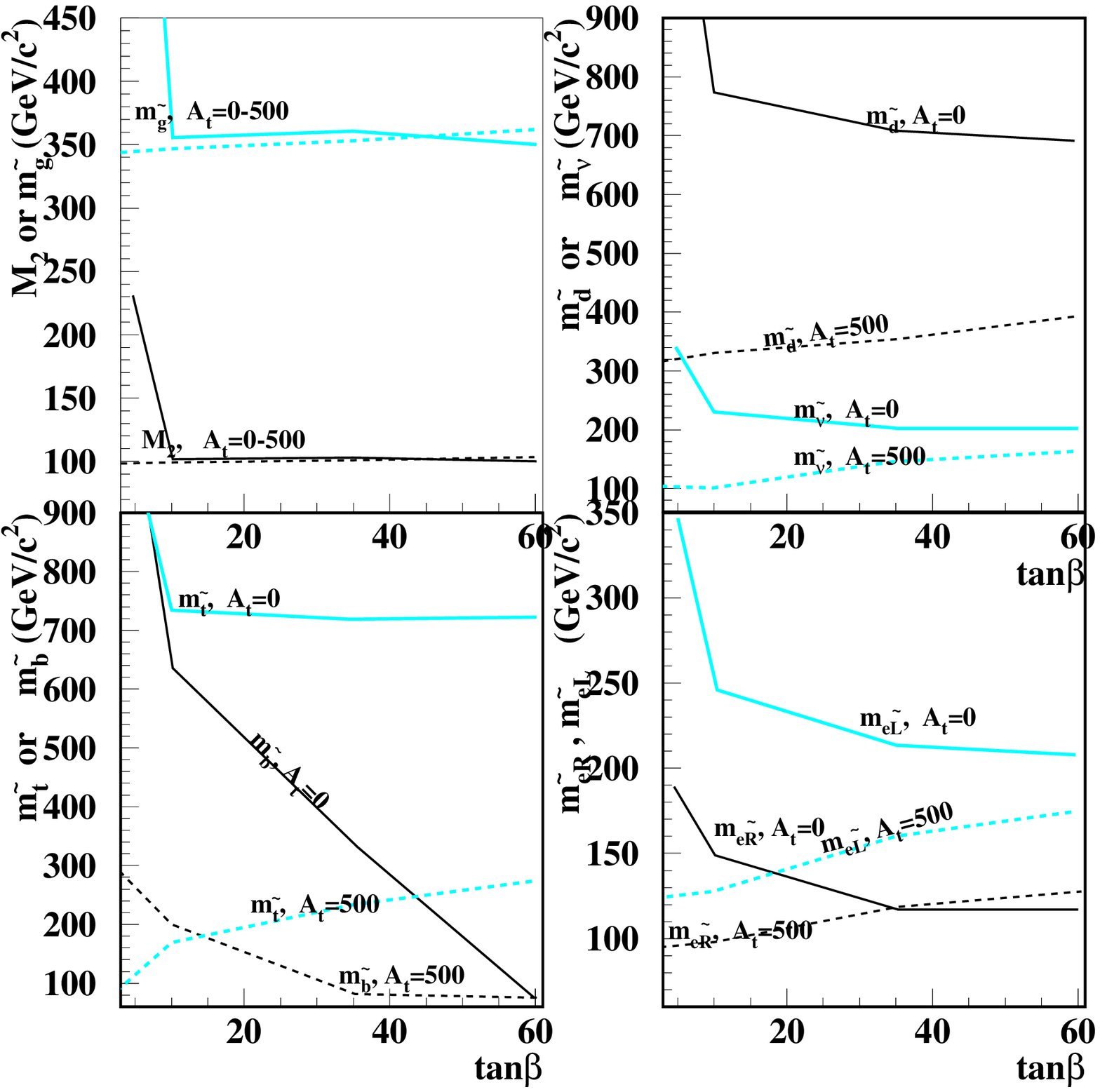}}
\caption[MSSM limits in ($\mu$,$M_2$) plane]{
From left to right, top to bottom:
Limits on $m_{\tilde{g}}$ and $M_2$, $m_{\tilde{d}}$ and \msnu, 
$m_{\stq_{1}}$ and $m_{\sbq_{1}}$,
$\mselr $, and 
$\msel\ $ as a function of \tanb, in the 
CMSSM with non-universal Higgs parameters.
Solid (dashed) lines  show limits for $A_t=0$ ($A_t=$ 500~\GeVcc  ).
}
\label{fig:m2lims}
\end{center}
\end{figure}

The Higgs mass depends  on $m_{\stq}$ via radiative corrections,  so Higgs 
searches can be  used to set a limit on $m_{\stq}$.
Figure \ref{fig:squark} shows the  
allowed range in  the ($m_{\stq_{1}}$, $m_{\sbq_{1}}$)
plane resulting from Higgs 
and SUSY searches, 
for several values of \tanb,  and for two example values
of $A_t$. Figure \ref{fig:m2lims} shows a lower limit
on $m_{\stq_{1}}$ and  $m_{\sbq_{1}}$ as a function of \tanb.
For $A_t =500$~\GeVcc, $m_{\stq_{1}} <$ 200~\GeVcc\ is allowed 
if $ \tanb <20$. If $A_t$=0~\GeVcc\ one gets
$m_{\stq_{1}}  \geqsim$ 700~\GeVcc\ for all studied values of
\tanb. 
A light sbottom ($\sim$ 100~\GeVcc) 
is allowed
for \tanb $\geqsim$~30, if $A_t$=500~\GeVcc. For $A_t=0$, one gets     
$m_{\sbq_{1}}> $ 200~\GeVcc\ for   \tanb$<$50.
If there is a Standard Model like Higgs boson lighter 
than 117~\GeVcc, an upper limit of $\sim$ 1000~\GeVcc\ exists
on   $m_{\sbq_{1}}$ and  $m_{\stq_{1}}$.

In the CMSSM with nUHP the mass of the lightest Higgs boson is not
directly related to $M_2$ value and (with all the caveats 
explained before) it is the chargino search which sets the limit 
on   $M_2$ for $A_t=500$, and for
$A_t=0$ if  $ \tanb >10$. For $A_t=0$, large $m_{SUSY}$ 
is nevertheless required to push
 $m_{\hn}$ up, which results in forcing large $m_0$ for small $M_2$.

Limits on $m_{\tilde{g}}$, $M_2$, $m_0$, $m_{\tilde{d}}$, \msnu, \mstq,
\msbq, 
 \msell\ and \mselr\  
as a function of \tanb\
 are shown on figure \ref{fig:m2lims}.
As expected, $m_{\tilde{g}}$ and  $m_{\tilde{d}} \geqsim$ 330~\GeVcc.
At high \tanb\
higher $m_0$ is required to avoid the stau becoming the 
 LSP. This is reflected
in a rise of   the limits on the slepton masses  and  $m_{\stq_{1}}$ 
with \tanb\
for $A_t=500$~\GeVcc\ (where the theoretical value of $m_{\hn}$ is above
the experimental limit even for relatively low $m_0$ values).
Limits of 100-150~\GeVcc\ are set for  
\mselr\ and \msnu, and 120-180~\GeVcc\
for \msell.\\

It should be noted that all limits discussed so far are not expected
to depend on the sign of $A_t$.\\

\noindent
\underline{Effect of $b \to s \gamma$ and Dark Matter
constraints}

As discussed in  section \ref{sec:other}, for $A_{t} \ge 0$,
$b \to s \gamma$ constraints
are not expected to affect any of the limits presented in this section.
If $A_t<0$, 
positive $\mu$  values are favoured, and for $\tanb>$20 and 
$A_t \simeq - 500$~\GeVcc\  
either $\stq_{1}$ or $\XPM{1}$ must be heavier than $ \sim $250~\GeVcc, or 
$m_{H^{\pm}} (m_A)$ must be of the order of 250(200)~\GeVcc. As Higgs and SUSY
searches impose $\mstq_{1}> 200$~\GeVcc\ for \tanb$>$20, $b \to s \gamma$
constraints do not tighten very much this limit.

In the  R-parity conserving scenario, the relict density of dark matter
can be used to set upper limits on sparticle masses, {\bf if the lightest
neutralino
is a gaugino}.
If one ignores  slightly fine-tuned possibilities of the
neutralino being very close in mass to \stauo\ or, (much less fine-tuned)  
to 
0.5~$m_{A,H} $ (resonant annihilation), 
the condition  $\Omega_{LSP} h <$~0.3  imposes 
$m_0 \leqsim $500~\GeVcc\ and $M_2 \leqsim 600$~\GeVcc\
for \tanb$<$50, and 
$m_0 \leqsim $150(200)~\GeVcc\ and $M_2 \leqsim 300$~\GeVcc\
for \tanb$<$10(20). This implies an upper limit on the LSP 
mass of 150-300~\GeVcc\ depending on \tanb,  
on \mselr\  of 200-600~\GeVcc, 
on \msell\  of 300-500~\GeVcc, and 
on $m_{\tilde{g}}$ of 1000-2000~\GeVcc. For \tanb$<10$ a mixing-dependent
upper limit 
of 800-1100~\GeVcc\ on the masses of $\stq_{1}$ and $\sbq_{1}$ is set.

\subsection{Limits in the mSUGRA scenario}

Limits on the mSUGRA model for $A_0=0$  were discussed
in detail in \cite{ellishigtb,roszk}. The Higgs search plays a major r\^ole
in setting these limits, and the value of $m_{\hn}$ depends crucially 
on $A_t \simeq 0.25 A_0 - 2m_{1/2}$, as it was
noted in for example \cite{wimheavy}. Here, $A_0$ values
in the range of $ (-500, 500$~\GeVcc) are studied.  The
dependence of the results on the accuracy of the Higgs
mass calculations is discussed in the following.  The ISASUGRA 
\cite{isasugra} model was used to calculate the sparticle spectrum  
and the values of the MSSM parameters at the EW scale, but the 
radiative corrections of ref. \cite{radcor} to chargino and
neutralino masses were implemented.
The calculations of $m_{\hn}$ of 
ref. \cite{carwig} were used, which give  $m_{\hn}$
typically 2-3~\GeVcc\ higher than in the ISASUGRA model.

\begin{figure}
\begin{center}
\vskip 0.5 cm
\mbox{\epsfysize=14.0cm\epsfxsize=14cm\epsffile{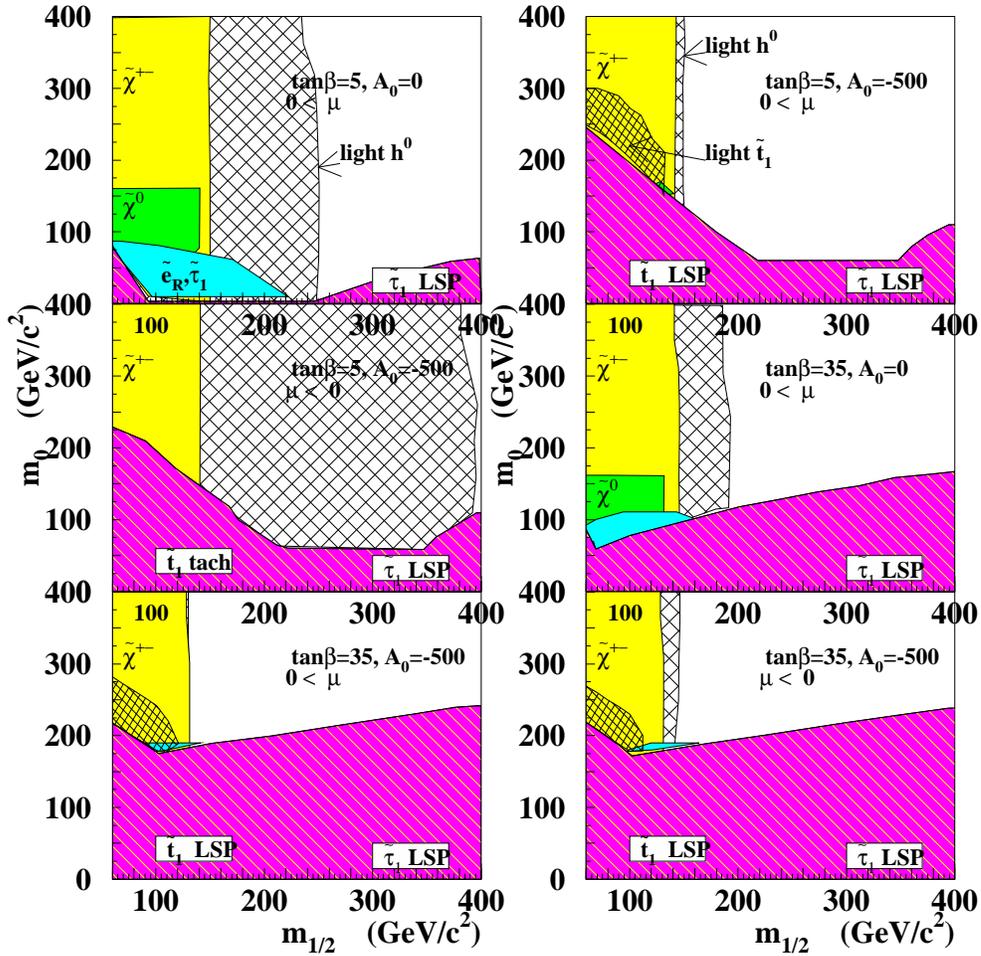}}
\caption[MSSM limits in ($\mu$,$M_2$) plane]{
Exclusion regions in the mSUGRA scenario from Higgs and SUSY searches
at LEP and stop searches at LEP and Tevatron Run I, for two values of \tanb\
and $A_0$. Light shaded vertical bands are excluded by chargino searches,
cross-hatched vertical bands are excluded by Higgs searches, 
and fine cross-hatched
areas for plots with $A_0=-500$~\GeVcc\ are excluded by stop searches at
LEP and Tevatron. Searches for neutralinos, \selr\ and \stauo\ (marked with
intermediate shading) complement
chargino searches for low $m_0$ in R-parity conserving scenarios.
In most  parity violating scenarios, a similar  vertical band is excluded
by chargino searches, but ``holes'' for small $m_0$ are absent. 
Dark hatched shading shows regions where either the \stauo\ or the $\stq_{1}$
is the LSP, or the $\stq_{1}$ is tachyonic.
}
\label{fig:planehi}
\end{center}
\end{figure}

To illustrate the effects of various searches, the corresponding exclusions
in the $m_0$ and $m_{1/2}$ plane are plotted on figure \ref{fig:planehi}
for two values of \tanb, $A_0$
and the sign of $\mu$. The value of $m_{top}=174.3$~\GeVcc\
was used. Chargino searches set
a limit on $m_{1/2}$ that is
 nearly independent $m_0$ and $A_0$.
Searches for
neutralinos, selectrons and staus help to cover holes in the chargino  
search at low $m_0$ arising in R-parity conserving scenarios.
Searches for the Higgs boson set a limit on $m_{1/2}$ which
depends very strongly on $A_0$ and $m_{top}$. It should be noted that
just a 1~\GeVcc\ change in the calculated value of the  \hn\ mass  can move   
the limit on $m_{1/2}$ 
set by the Higgs boson searches  by 30-150~\GeVcc\ 
 (\footnote{ The Higgs mass calculations of \cite{carwig} used in this
paper give values 1-3 \GeVcc\ higher than FeynHiggs used
for example in \cite{benchellis}. Limits presented here are
thus somewhat more conservative. With this conservative
Higgs mass calculation $m_0=0$ is still allowed
for $\tanb \simeq 5$ (see
figure \ref{fig:planehi}), as remarked in \protect{\cite{ellisscalar}}}.) 
depending on \tanb ! For $A_0 =-500$~\GeVcc\ and $\mu>0$, 
Higgs searches exclude
$m_{1/2}$ values just about 25~\GeVcc\ higher than these excluded by 
chargino searches already at \tanb$>$5.

As remarked in  section \ref{sec:other}, in mSUGRA $\mu>0$ is favoured due to
strong constraints from $b \to s \gamma$ on $\mu<0$.

\begin{figure}
\begin{center}
\vskip 0.5 cm
\mbox{\epsfysize=14.0cm\epsfxsize=14cm\epsffile{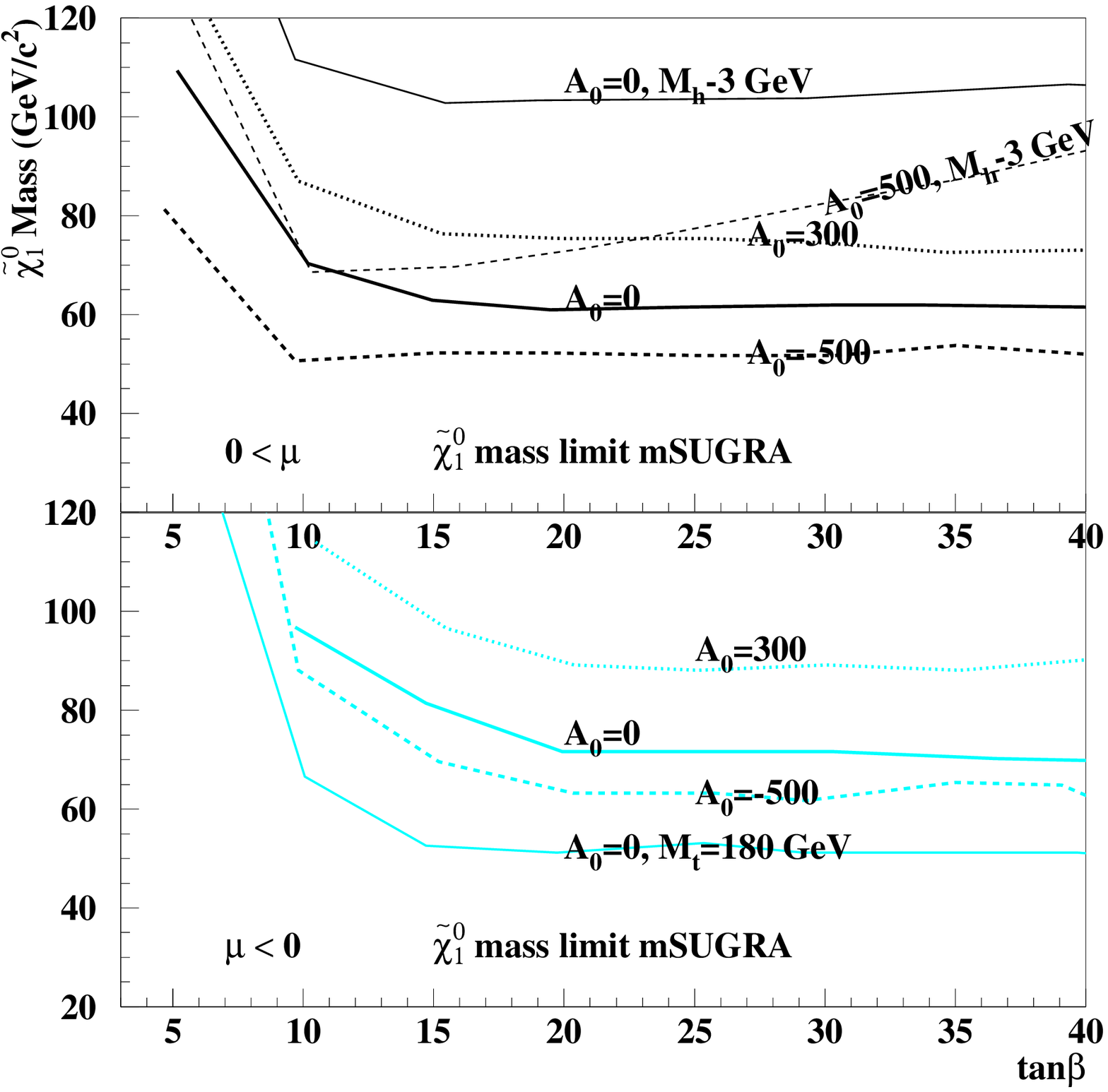}}
\caption[MSSM limits in ($\mu$,$M_2$) plane]{
The lower limit at 95~\% confidence level on the mass of the lightest
neutralino, \XN{1}, in mSUGRA. The limits for positive $\mu$
(upper plot) and negative $\mu$ (lower plot) are 
shown.
The dashed, solid, and dotted curves shows limits obtained for 
$A_0$~=-500~\GeVcc, $A_0$~=0  and $A_0$~=300~\GeVcc, respectively.
The thin solid curve on the lower plot
shows the limit for $A_0=0$ and
$m_{top}=180.0$~\GeVcc. $m_{top}=174.3$ \GeVcc\ 
for all other curves.  The LSP limit degrades in this case down
to the one set by chargino searches for $\tanb>15$.
Thin solid (dashed) curves on the upper plot  show the limit  obtained for 
 $A_0$~=0~\GeVcc\ ($A_0$~=-500~\GeVcc) with the  calculated $m_{\hn}$ value
lowered by 3~\GeVcc\ (this corresponds to a  rise in the experimental
limit).
These limits result primarily
from chargino and Higgs searches and are also valid in R-parity
violating scenarios as long as a kinematic limit on the chargino
mass can be set. 
If $b \to s \gamma$ and dark matter (R-parity conserving scenario)
constraints were used, the lower limit on the LSP mass would rise to
$\sim 110~$\GeVcc\ for \tanb$>$20.
}
\label{fig:lspmsugra}
\end{center}
\end{figure}

Excluded regions in $m_{1/2}$ can be translated  into limits on
$\MXN{1}$ and $\MXC{1}$. Limits on $\MXN{1}$ are illustrated
on figure \ref{fig:lspmsugra} for several values of $A_0$
and $m_{top}$. $\MXC{1}$ is close to 2$\MXN{1}$.
For $A_0=-500$~\GeVcc\ the Higgs search does not
contribute for $\tanb>10$ and limit is given by the SUSY searches.
However, Higgs searches become more constraining for positive
$A_0$ values, as long as they lead  to
an appreciable  decrease of the absolute
value of $X_t
\simeq 0.25 A_0 -2m_{1/2}(1+cot\beta)$ 
 for $m_{1/2}$ values where the chargino limit
is no longer effective ($\geqsim 150$~\GeVcc).
For $A_0$ and \tanb\ values where the Higgs search contributes,
the LSP limit  can change by 50~\GeVcc\ for a 3~\GeVcc\ change of the
Higgs mass. 
The $\MXN{1}$ limit presented here is in agreement with the
conservative ``prediction'' of the final LSP mass limit from \cite{ellislight}
and it is more conservative than the result presented in \cite{ellishigtb},
probably due to a more conservative Higgs mass calculation.

\begin{figure}
\begin{center}
\vskip 0.5 cm
\mbox{\epsfysize=14.0cm\epsfxsize=14cm\epsffile{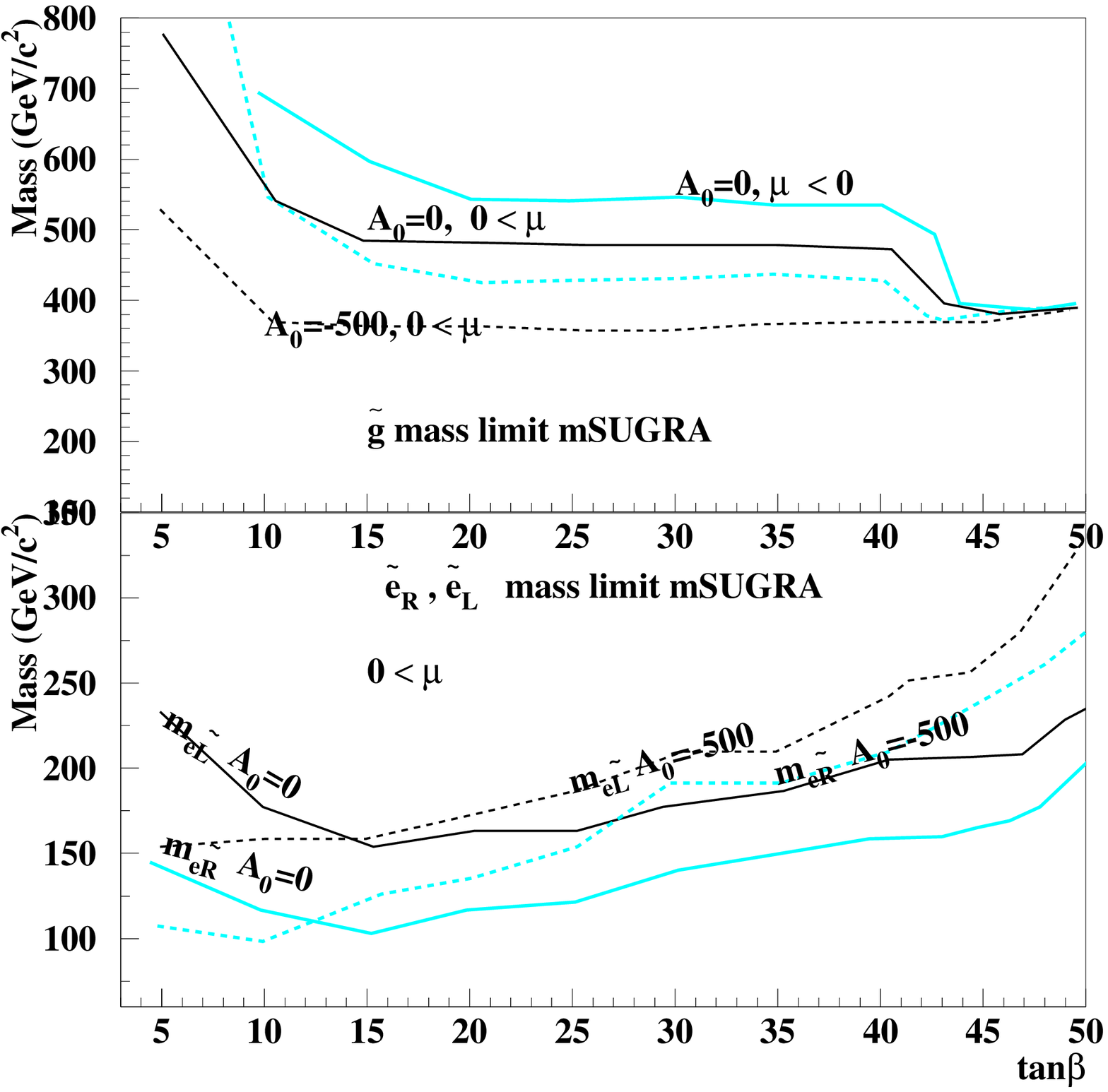}}
\caption[MSSM limits in ($\mu$,$M_2$) plane]{
The lower limit at 95~\% confidence level on the masses $\tilde{g}$,
\selr\ and \sell,  in mSUGRA. The limits for positive $\mu$
(negative $\mu$) are 
shown in thin dark (thick light) lines.
The dashed (solid) curve shows limits obtained for 
$A_0$~=-500~\GeVcc\ ($A_0$~=0~\GeVcc). See text for more
explanations.
These limits result primarily
from chargino, \selr\ and Higgs searches and are also valid in R-parity
violating scenario, as long as the kinematic limit on the chargino
mass can be set. In R-parity violating
scenarios regions, where the \selr\ limit
is effective, and which are not
covered by the Higgs search, are covered  either by
chargino searches or by stop searches, or by the requirement of a  
non-tachyonic stop.
These limits can be tighten for \tanb$>$ 20, 
if $b \to s \gamma$ and dark matter (R-parity conserving scenario)
constraints are used (see text).
}
\label{fig:glussug}
\end{center}
\end{figure}

Figure \ref{fig:glussug} illustrates limits on 
$m_{\tilde{g}}$, \msel\ and \mselr.
These limits are close to the ones obtained in 
the CMSSM scenario. For $A_0=-500$~\GeVcc,
the Higgs limit is not constraining 
for $\tanb>10$ and $\mu>0$. At higher \tanb\
there is a limit on $m_0$ set by the  requirement that \stauo\ should be 
heavier than
\XN{1}, and limits on the sleptons rise. 
The Higgs constraint degrades for $\tanb>40$ due to a light 
$A$ boson being allowed, which
results in a decrease of the experimental sensitivity (see section 
\ref{sec:higgs}).

\begin{figure}
\begin{center}
\vskip -0.5 cm
\mbox{\epsfysize=13.0cm\epsfxsize=13cm\epsffile{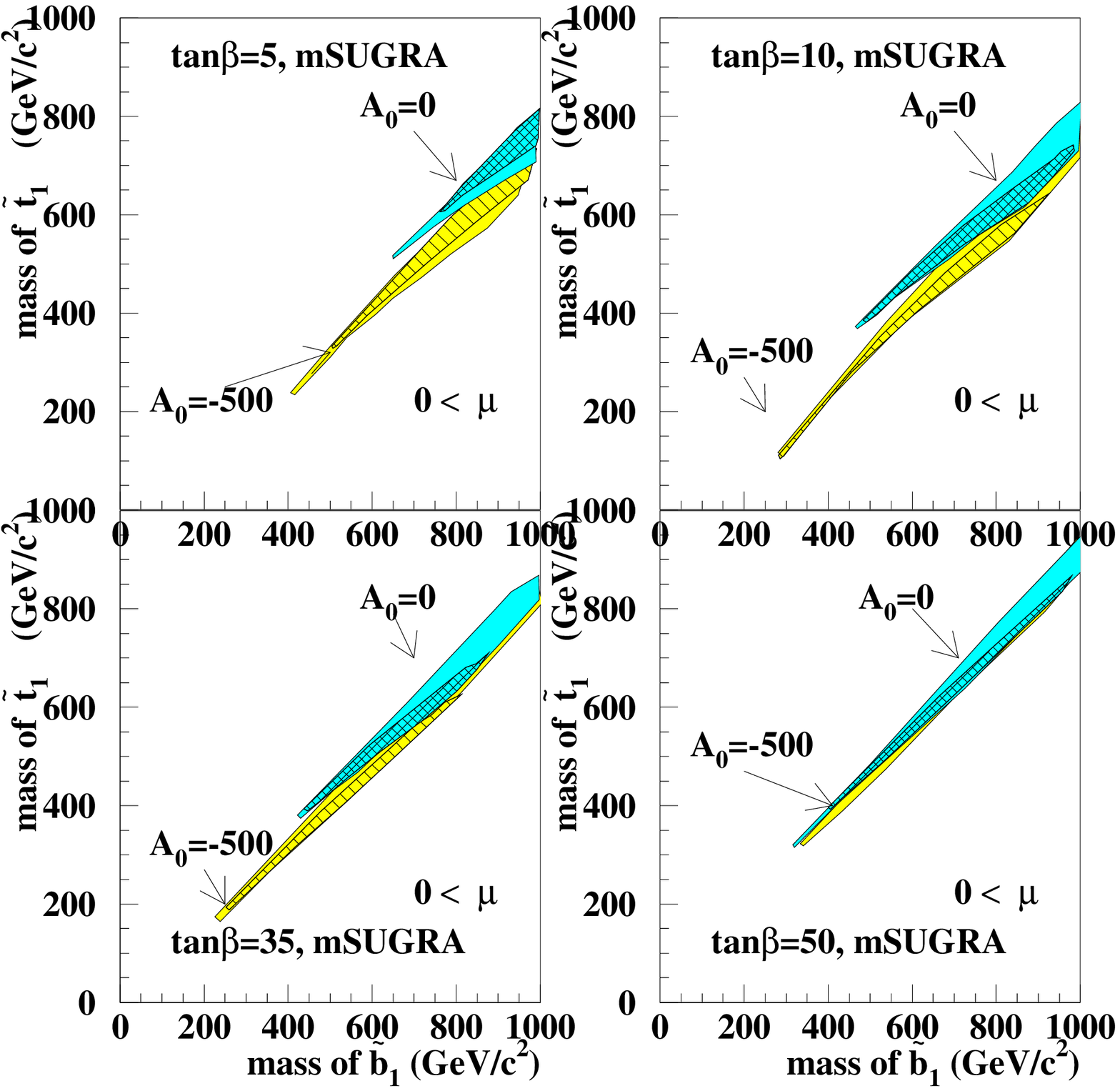}}
\caption[MSSM limits in ($\mu$,$M_2$) plane]{
Allowed range in  ($m_{\stq_{1}}$, $m_{\sbq_{1}}$)
plane, at 95~\% confidence level
for several values of \tanb, resulting from Higgs 
and SUSY searches.  The darker (lighter)  region is
for $A_0=0$ ($A_0$=-500~\GeVcc). The hatched (cross-hatched)
region on the lighter (darker) shading is allowed for
0.03~pb $\leqsim$ $\sigma(\hn Z)$BR($\hn \to b \bar{b}$) $\leqsim$ 0.07~pb.
Only the region  $\mu>0$ is shown, as it is less restricted.
If  $ b \to s \gamma$ constraints from \cite{benchellis} are applied, the
lower limit on the mass of the  $\stq_{1}$ ($\sbq_{1}$) grows to 
300~\GeVcc\ (450~\GeVcc) for \tanb$>$ 20 (see text). 
}
\label{fig:squarksug}
\end{center}
\end{figure}

Figure \ref{fig:squarksug} shows the  allowed 
range in  the  ($m_{\stq_{1}}$, $m_{\sbq_{1}}$)
plane resulting from Higgs 
and SUSY searches in mSUGRA, 
for several values of \tanb,  and for two example values
of $A_0$. 
For $A_0 =-500$~\GeVcc, $m_{\stq_{1}}$ $<$ 200~\GeVcc\ is marginally
allowed  
for $10< \tanb < 20$.  $m_{\stq_{1}}$  $\geqsim$ 300~\GeVcc\ for
 $A_0$=0. 
The lightest sbottom is heavier than
200~\GeVcc\ for all the range 
of $A_0$ and \tanb\ studied.
For $\mu>0$, which is less restricted by $ b \to s \gamma$
constraints, 
the possible ``evidence'' for the light \hn\
sets an upper limit of close to 1000~\GeVcc\ on the masses of 
 $\stq_{1}$ and $\sbq_{1}$.

Constraints resulting from the 
$ b \to s \gamma$ measurement can tighten the above
limits on sleptons and squarks at $\tanb>20$, while limits on $\MXN{1}$,
$\MXC{1}$ and $m_{\tilde{g}}$ are not going to be affected 
as $ b \to s \gamma$
does not constrain $m_{1/2}$ for high $m_0$ (heavy squarks).  
However, if one applies both    $ b \to s \gamma$ limits
and an upper limit on $m_0$ for small $m_{1/2}$ resulting
from the upper limit on the relict density (R-parity conserving scenario),
 lower limits on $\MXN{1}(\MXC{1}) \geqsim 110(220) $~\GeVcc\ 
and $m_{\tilde{g}} \geqsim 700$~\GeVcc\ will result for \tanb$>$20.
Results of \cite{roszk} and \cite{benchellis} suggest that, at
least for $A_0=0$, there is little 
room for spectra which are consistent both with $ b \to s \gamma$ and
an upper limit on the relic density for \tanb$>$35, unless resonant neutralino
annihilation and co-annihilation are taken into account.

If  $ b \to s \gamma$ constraints from \cite{benchellis} are applied, the
lower limit on the mass of the  $\stq_{1}$ ($\sbq_{1}$) grows to 
300~\GeVcc\ (450~\GeVcc) for \tanb$>$ 20. 
However, 
these constraints have to be recalculated
for $A_0$ values other than 0. In particular,
large positive values of $A_0$ (such that
$A_t$  is positive at $m_{1/2} \sim 150$~\GeVcc) should be considered,
as they weaken $ b \to s \gamma$ constraints \cite{carenam}.
Moreover mass constraints above cannot be interpretted a
 95\% confidence limits with the present treatment of the
experimental and theoretical errors on  $ b \to s \gamma$.

It is interesting to note that both 
the possible ``evidence'' for the light \hn\
and ``naively'' applied dark matter constraints (ignoring 
the neutralino-stau co-annihilation, and the resonant neutralino annihilation) 
imply an upper limit of close to 1000~\GeVcc\ on the masses of 
 $\stq_{1}$ and $\sbq_{1}$, and an upper limit of 150-250~\GeVcc\
 on the LSP mass.

\section{Searches at the Tevatron Run II and a summary
of LEP and Run I limits}
\label{sec:tevatron}. 

Although finding a light \hn\ at the Tevatron
would constitute an interesting suggestion that   the MSSM is the
right extension of the Standard Model, 
only  the observation   of the sparticles 
can establish its validity.

Several channels of sparticle production in R-parity
conserving scenario are considered for  Run II
of the Tevatron, among them the production of $\XN{1} \XN{2}$,  
$\XPM{1} \XN{2}$, $\stq_{1} \stq_{1}$, $\sbq_{1} \sbq_{1}$,
gluino and squark production are the most important ones
\cite{sugtev,numerots}.
An approximate reach of  Run II with 25 (2) $fb^{-1}$ in the masses 
of the above sparticles  is
150 (110)~\GeVcc\ for   $\MXC{1}$, 75 (55)~\GeVcc\ for  $\MXN{1}$
and 450 (330)~\GeVcc\ for the gluino mass, provided $m_0 \leqsim 200$~\GeVcc.
The stop, $\stq_{1}$, is observable up to 260(180)~\GeVcc\ and the $\sbq_{1}$
up to 280(210)~\GeVcc. \\

In R-parity violating scenarios, where the LSP
decays to the final states containing leptons (via leptonic  $\lambda$  and
leptonic-hadronic $\lambda'$ coupling) higher reach 
was reported for the sparticle pair-production \cite{paritytev}.
This is primarily due to the better visibility of the final states.
Using all the SUSY production channels and assuming mSUGRA mass relations,
gluino with mass below 500-600~\GeVcc\ can be excluded at 
Run IIa ($2~fb^{-1}$). Searches for single sparticle production can have  a
higher
kinematic reach, but  production cross-sections depend directly on the
value of the R-parity violating coupling involved.

LEP and Tevatron Run I searches place relevant limits
on the masses of the  all above particles.

In the constrained MSSM with non-universal Higgs parameters,
the lightest chargino, the lightest neutralino and the gluino could 
be in the range 
of the Tevatron Run II for \tanb$>$3, but only in the high luminosity scenario.
For large mixing in the stop sector, the stop lighter than 200~\GeVcc\
is allowed for  $3 \leqsim \tanb \leqsim 20-30$, 
and $\sbq_{1}$ lighter than 200~\GeVcc\ is allowed 
for \tanb$ \geqsim $10. These squarks could thus be   
observable in Tevatron Run IIa (2$fb^{-1}$). 
However, for small mixing they
are above the range of the Tevatron.

Light stau $m_{\stauo} \geqsim 87$ is not excluded by LEP II constraints,
and could be perhaps observable at the Tevatron.

In the mSUGRA scenario, gauginos could be observed at the 
Tevatron (Run IIb) if 
$A_0$ is large and \tanb$\geqsim$ 7.
Gluino can be lighter than 500-600~\GeVcc\ for $\mu>0$,\tanb$\geqsim$ 7, thus 
the R-parity violating SUSY can be observed already
at the Run IIa, if the LSP decays give electrons or
muons. 
The stop, $\stq_{1}$, can be lighter than 200~\GeVcc\  and
thus in the range of Run IIa , for large $A_0$ and
$ 10 \leqsim \tanb \leqsim 30$.
The sbottom
can be somewhat lighter than 300~\GeVcc\ for    large $A_0$ and 
$ 10 \leqsim \tanb  \leqsim 40$,
thus perhaps in the range of  Run IIb.
However, if $b \to s \gamma$ constraints are used, the stop and 
sbottom are beyond the
reach of the Tevatron for $\tanb \geqsim 20$. If both  $b \to s \gamma$ and cosmological
constraints are used (R-parity conservation), 
gauginos (chargino, neutralino, gluino) 
are beyond the
reach of the Tevatron for  $\tanb \geqsim 20$. 
For $ 7 \leqsim \tanb \leqsim  20$  it is
still possible to observe gauginos in Run IIb 
(for large $A_0$) and the possibility of their
discovery  is enhanced  due to 
the upper limit set on $m_0$ by the relict density of the LSP's.  
\\

Searches for the stop and the sbottom  thus seem to be the most promising
SUSY disovery channels at   Run IIa. Below a few examples  of non-excluded
sets of CMSSM and mSUGRA parameters corresponding to 
the light stop or sbottom are given (parameters in 
\GeVcc\ whenever appropriate):

MSSM: \tanb=10, $m_0=154$, $\mu=-1200$, $M_2=106$, $A_t=500,
A_b=A_{\tau}=0$, $m_A=1000$. In this point \mstq=198  and
\msbq=246. The stop decays to \XPM{1} b, and $\XPM{1} \to \stauo \nu$,
$m_{\stauo}$=100~\GeVcc.

MSSM: \tanb=10, $m_0=80$, $\mu=300$, $M_2=133$, $A_t=-800,  
A_b=-800, A_{\tau}=-495$, $m_A=1000$. In this point \mstq=194, and
\msbq=367. The stop decays to \XPM{1} b, and $\XPM{1} \to \stauo \nu$,
$m_{\stauo}$=96~\GeVcc.

MSSM: \tanb=35, $m_0=300$, $\mu=1519$, $M_2=160$, $A_t=1100, 
A_b=A_{\tau}=0$, $m_A=1000$. In this point \mstq=374  and
\msbq=177. The sbottom decays to \XN{1} b  (80\%), 
\XN{2} b  (20\%) and  $\XN{2} \to \stauo \tau$. $m_{\stauo}$=100~\GeVcc.

mSUGRA: \tanb=10, $m_0=80$, $m_{1/2}=160$, $A_0=-400$, $\mu>0$. 
In this point \mstq=176  and
\msbq=318.

mSUGRA: \tanb=15, $m_0=100$, $m_{1/2}=180$, $A_0=-500$, $\mu>0$. 
In this point \mstq=186  and
\msbq=309.

In many of these points 
\stauo\ is relatively light and final states with $\tau$'s 
are important.\\

However three-body decays $\stq_{1} \to b l \snu$ are not important 
for the stop detectable at the
Tevatron, due to the high limit on $\msnu$ set by LEP (see previous section).  
For the same reason, the decays 
$\stq_{1} \to b \sle \nu$ are only important if \sle=\stauo.\\
 
In summary, the stop and the sbotom searches are the most promising
discovery channels of gravity-mediated SUSY with R-parity conservation.
As also pointed out in \cite{djouadi} the final states containing
taus should be given a special attention.
For $\mu>0$ and tanb$\geqsim$ 7, 
R-parity violating SUSY can be observed already
in  Run IIa  if the LSP decays give electrons or
muons. \\

It is interesting to note that in the constrained
MSSM with   non-universal Higgs parameters,
the requirement that there is a Standard-Model-like
 \hn\ with $m_{\hn}<117$~\GeVcc\
sets an upper limit of around 
1~\TeVcc\ on the mass of the lightest stop. A similar
limit is set by the requirement 
that $\Omega_{LSP}h^2<$~0.3 if the lightest neutralino is
a gaugino  and the resonant annihilation
and resonant co-annihilation are ignored. 
The  mass region 
$50 \leqsim \MXN{1} \leqsim 300$~\GeVcc\ is preferred.
If the lightest neutralino is a higgsino, $\MXN{1}>80$~\GeVcc\ is set by 
the LEP
results, excluding a possibility of the light higgsino Dark Matter 
(see \cite{edsjo}).

Similar conclusions can be drawn in the mSUGRA scenario;
the results of \cite{roszk} and \cite{benchellis} 
suggest though that (for $A_0=0$ and \tanb$>$35) there is little 
room for spectra which are consistent both with $ b \to s \gamma$ and
with an upper limit on the relic density, outside
the regions allowed by  the resonant neutralino
annihilation and the resonant neutralino-stau co-annihilation.
However, 
these constraints have to be recalculated
for $A_0$ values other than 0. In particular,
large positive values of $A_0$ such that
$A_t$  is positive at $m_{1/2} \geqsim 150$~\GeVcc, should be considered,
as they weaken $ b \to s \gamma$ constraints \cite{carenam}. They
can allow for lighter sparticles
at large \tanb, where the LEP Higgs
mass limit is less effective in terms of constraints on $m_{1/2}$.
Moreover 
the treatment of the
experimental and theoretical errors on  $ b \to s \gamma$  has to be improved
to enable the statistical interpretation of
the resulting constraints.

\subsection*{Acknowledgements}
\vskip 3 mm
I would like to thank Wilbur Venus and Janusz Rosiek for reading
the manuscript and many useful comments.
I would like to thank my DELPHI colleagues for many discussions on this
subjects.



\end{document}